\documentclass[aps,prb,twocolumn,amsmath,amssymb,nofootinbib,eqsecnum,superscriptaddress,floatfix]{revtex4}

\usepackage{amsmath}
\usepackage{amssymb}
\usepackage{amsthm}
\usepackage[dvips]{color}
\usepackage{bm} 
\usepackage[hypertex]{hyperref}
\usepackage{longtable}
\usepackage{ulem}   
\newcommand{\bs}[1]{{\boldsymbol{#1}}}

\begin{document}

\title{ 
Magnetic translation algebra with or without magnetic field\\
in the continuum or on arbitrary Bravais lattices in any dimension
      }

\date{\today}

\author{Claudio Chamon} 
\affiliation{
Physics Department,
Boston University, 
Boston, Massachusetts 02215, USA
            } 

\author{Christopher Mudry} 
\affiliation{
Condensed Matter Theory Group, 
Paul Scherrer Institute, CH-5232 Villigen PSI,
Switzerland
            } 

\begin{abstract}
The magnetic translation algebra plays an important role
in the quantum Hall effect. Murthy and Shankar, 
arXiv:1207.2133, have shown how to realize this algebra 
using fermionic bilinears defined
on a two-dimensional square lattice. We show that, 
in any dimension $d$, 
it is always possible to close the magnetic translation algebra 
using fermionic bilinears, whether in the continuum or on the lattice. 
We also show that these generators are complete in even, but not odd, 
dimensions, in the sense that any
fermionic Hamiltonian in even dimensions
that conserves particle number can be represented in terms 
of the generators of this algebra,
whether or not time-reversal symmetry is broken. 
As an example, we reproduce the $f$-sum rule of interacting
electrons at vanishing magnetic field using this representation.
We also show that interactions
can significantly change the bare bandwidth
of lattice Hamiltonians when represented in terms of the
generators of the magnetic translation algebra.
\end{abstract}

\maketitle

\section{
Introduction
        }

The two coordinates of an electron in the plane orthogonal to
a uniform magnetic field 
$B\,\bs{e}$ pointing along the direction $\bs{e}$ 
do not commute. There follows
the $U(1)$ algebra
\begin{equation}
[t(\bs{a}),t(\bs{b})]=
\mathrm{i}\,
\sin
\left(\frac{(\bs{a}\wedge\bs{b})\cdot\bs{e}}{2}\right)\,
t(\bs{a}+\bs{b})
\label{eq: def magnetic algebra}
\end{equation}
obeyed by the triplet of generators 
$t(\bs{a})$,
$t(\bs{b})$,
and
$t(\bs{a}+\bs{b})$
for any pair $\bs{a}$ and $\bs{b}$ 
of vectors orthogonal to $\bs{e}$,
which is called the magnetic translation algebra in this context.%
~\cite{Zak64}
The magnetic translation algebra
can be used to derive the transverse conductivity of 
the integer quantum Hall effect (IQHE).
It has also been used by Girvin, MacDonald, and Platzman
in Ref.~\onlinecite{Girvin85}
to place a variational estimate on the excitation gap 
for the fractional quantum Hall effect (FQHE), 
following closely the approach of
Feynman and Bijl in their study of excitations in $^4$He.%
~\cite{Feynman72}

Hamiltonians defined on two-dimensional lattices with topologically
nontrivial bands can also display quantum Hall physics. The IQHE can
occur in band insulators when the Bloch bands have a nonvanishing
Chern number, as shown by Haldane.%
~\cite{Haldane88} The FQHE effect requires strong electronic
correlations. This is possible if the Chern bands are 
sufficiently narrow (or even flat).%
~\cite{Neupert11a,Tang11,Sun11} Whether flat Chern bands can sustain 
a FQHE or not is a matter of energetics. Exact diagonalization studies of
fractionally filled Chern bands with added short-range interactions
are consistent with a correlated liquid ground state 
supporting a FQHE for certain filling fractions.%
~\cite{Neupert11a,Sheng11,Wang11a,Regnault11a,Neupert11b,Bernevig11,Venderbos12,Wang12a,Bernevig12a,Bernevig12b,Wang12b,Liu12,Grushin12,Lauchli12,Sterdyniak12}
Such topological correlated states on the lattice are now known as
fractional Chern insulators (FCI).

In an effort to draw a bridge between the case when the FQHE
is realized in the continuum in contrast to the case when
it is realized in a FCI,
Parameswaran, Roy, and Sondhi in Ref.~\onlinecite{Parameswaran12} 
have pioneered an algebraic approach
to FCIs by deriving the algebra obeyed by the density operators
projected to the partially filled band.%
~\cite{Bernevig11,Goerbig,Neupert12b,Estienne12}
They found that the algebra~(\ref{eq: def magnetic algebra}) 
emerges to leading order in a gradient expansion.
Remarkably, Murthy and Shankar have (i) constructed 
in Ref.~\onlinecite{Murthy12}
a coherent superposition of the projected 
density operator that closes the $U(1)$ algebra%
~(\ref{eq: def magnetic algebra})
on the square lattice and (ii) represented
any Hamiltonian that commutes with the number operator
and describes the competition between
the electronic hopping and the electronic interaction
in terms of these generators.%
~\cite{Murthy11}

In this paper, we are going to generalize the
results by Murthy and Shankar as follows.
We shall represent the $U(1)$ algebra%
~(\ref{eq: def magnetic algebra})
in terms of coherent superpositions of electron-hole
pairs in arbitrary dimensions
both in the continuum and for Bravais lattices.
We shall then show that these generators
provide a complete basis for the linear space 
of operators spanned by charge-neutral fermion bilinears
provided the Bravais lattice, or its embedding
space in the continuum limit, is even dimensional.
For odd dimensions, the generators
of the $U(1)$ algebra%
~(\ref{eq: def magnetic algebra})
form an incomplete basis of the 
space of operators
spanned by charge-neutral fermion bilinears.

We first treat the case of Hamiltonians acting on 
wave functions supported in the continuum 
for pedagogical reasons
in Sec.~\ref{sec: continuum}.
After this warm-up,
we turn our attention to Hamiltonians acting on 
wave functions supported on Bravais lattices in Sec.%
~\ref{sec: lattice}. 
Sections~\ref{sec: continuum} and 
\ref{sec: lattice}
constitute the main results of this paper.

As a sanity check, we verify 
that the $f$-sum rule is
obeyed if one represents the electronic density operator
in terms of the particle-hole generators of the algebra%
~(\ref{eq: def magnetic algebra}) 
in Sec.~\ref{subsec: f sum rule}.
This exercise also suggests caution when performing
uncontrolled approximations using the magnetic algebra,
for such uncontrolled approximations could predict
effects associated to a spurious breaking of time-reversal symmetry.

In Sec.~\ref{subsec: Normal ordering and the bare band width},
we explain how,
when represented in terms of these
generators of the algebra%
~(\ref{eq: def magnetic algebra}),
interactions induce one-body terms that
can significantly change the bare bandwidth
of lattice Hamiltonians. The same effect 
in the FQHE requires the addition of a strong one-body
perturbation to a Landau band, one that is of the order of the FQHE gap. 
Thus, whereas the FQHE is a strong-coupling problem,
the FCI in a flat band is more like a problem at intermediate coupling.
This result explains why in Ref.%
~\onlinecite{Grushin12}
a FCI with a Chern number of 2 was more stable if the bare dispersion 
was not flat rather than flat, for the bare and induced one-body terms
can act to neutralize each other.

\section{
The case of the continuum
        }
\label{sec: continuum}

We define the fermionic Fock space $\mathfrak{F}$ with the help
of the algebra
\begin{equation}
\begin{split}
&
\left\{
\hat{c}       (\bs{k} ),
\hat{c}^{\dag}(\bs{k}')
\right\}=
\delta(\bs{k}-\bs{k}'),
\\
&
\left\{
\hat{c}(\bs{k} ),
\hat{c}(\bs{k}')
\right\}=
\left\{
\hat{c}^{\dag}(\bs{k} ),
\hat{c}^{\dag}(\bs{k}')
\right\}=0
\end{split}
\label{eq: def fermion algebra continuum}
\end{equation}
for any pair of momenta $\bs{k},\bs{k}'\in\mathbb{R}^{d}$.
Without loss of generality,
we ignore any internal degrees of freedom such as
the spin quantum numbers since we are after the
$U(1)$ algebra~(\ref{eq: def magnetic algebra}).

The linear space of fermionic bilinears that we
study is spanned by the basis
\begin{subequations}
\begin{equation}
\hat{T}(\bs{q}^{\ }_{1},\bs{q}^{\ }_{2}):=
\hat{c}^{\dag}(\bs{q}^{\ }_{1})\,
\hat{c}(\bs{q}^{\ }_{2}),
\label{eq: def T's continuum}
\end{equation}
which obeys the algebra
\begin{equation}
\begin{split}
\left[
\hat{T}(\bs{q}^{\,}_{1},\bs{q}^{\,}_{2}),
\hat{T}(\bs{q}^{\prime}_{1},\bs{q}^{\prime}_{2})
\right]=&\,
\delta(\bs{q}^{\,}_{2}-\bs{q}^{\prime}_{1})\,
\hat{T}(\bs{q}^{\,}_{1},\bs{q}^{\prime}_{2})
\\
&\,
-
\delta(\bs{q}^{\,}_{1}-\bs{q}^{\prime}_{2})\,
\hat{T}(\bs{q}^{\prime}_{1},\bs{q}^{\,}_{2})
\end{split}
\label{eq: algebra T's continuum}
\end{equation}
\end{subequations}
for any quadruple 
$\bs{q}^{\,}_{1}$,
$\bs{q}^{\,}_{2}$,
$\bs{q}^{\prime}_{1}$,
and
$\bs{q}^{\prime}_{2}$
from $\mathbb{R}^{d}$.

For any momentum $\bs{q}\in\mathbb{R}^{d}$ 
and for any function
$
f:\mathbb{R}^{d}\times\mathbb{R}^{d}
\longrightarrow\mathbb{C}
$,
define the coherent superposition
\begin{subequations}
\begin{equation}
\hat{\varrho}^{f}(\bs{q}):=
\int\limits_{\bs{p}}
f(\bs{q},\bs{p})\,
\hat{c}^{\dag}(\bs{q}+\bs{p})\,
\hat{c}(\bs{p}).
\label{eq: def rho f continuum}
\end{equation}
There follows the algebra
\begin{widetext}
\begin{equation}
\begin{split}
\left[
\hat{\varrho}^{f }(\bs{q} ),
\hat{\varrho}^{f'}(\bs{q}')
\right]=&\,
\int\limits_{\bs{p}}
\Big[
f(\bs{q},\bs{q}'+\bs{p})\,
f^{\prime}(\bs{q}',\bs{p})
-
(
\bs{q}\leftrightarrow\bs{q}'
\hbox{ and }
f\leftrightarrow f'
)
\Big]\,
\hat{c}^{\dag}(\bs{q}+\bs{q}'+\bs{p})\,
\hat{c}^{\,  }(\bs{p})
\label{eq: commutator two rho f's continuum}
\end{split}
\end{equation}
\end{widetext}
\end{subequations}
for any pair of momenta $\bs{q}$ and $\bs{q}'$
and for any pair of functions $f$ and $f'$.

The choice $f(\bs{q},\bs{p})=1$
for any pair of momenta
$\bs{q}$ and $\bs{p}$ from $\mathbb{R}^{d}$
defines the momentum representation of the 
local density operator
\begin{subequations}
\begin{equation}
\hat{\rho}(\bs{q}):=
\int\limits_{\bs{p}}
\hat{c}^{\dag}(\bs{q}+\bs{p})\,
\hat{c}(\bs{p}).
\label{eq: def rho continuum}
\end{equation}
Any pair thereof commutes as
\begin{equation}
\left[
\hat{\rho}(\bs{q} ),
\hat{\rho}(\bs{q}')
\right]=0.
\label{eq: algebra densities continuum}
\end{equation}
\end{subequations}

Another choice of the function $f$ is made with
the family 
\begin{subequations}
\label{eq: def rgo G q continuum}
\begin{equation}
\hat{\varrho}(\bs{q};\bs{G}):=
\int\limits_{\bs{p}}\,
e^{
+\mathrm{i}\,\Phi(\bs{q},\bs{p};\bs{G})
  }\,
\hat{c}^{\dag}(\bs{q}+\bs{p})\,
\hat{c}^{\,}(\bs{p}),
\label{eq: def rgo G q continuum a}
\end{equation}
for any pair $\bs{q}$ and $\bs{G}$ from $\mathbb{R}^{d}$,
where 
\begin{equation}
\Phi(\bs{q},\bs{p};\bs{G}):=
(\bs{q}+\bs{G})\,\bs{\ast}\,\bs{p}
-
\frac{1}{2}
\bs{q}\,\bs{\ast}\,\bs{G}
\label{eq: def rgo G q continuum b}
\end{equation}
and the $\bs{\ast}$ product
\begin{equation}
\bs{a}\,\bs{\ast}\,\bs{b}=
-
\bs{b}\,\bs{\ast}\,\bs{a}\equiv
\sum_{i,j=1}^{d}
a^{\,}_{i}\,
M^{(\bs{\ast})}_{ij}
b^{\,}_{j}
\label{eq: def rgo G q continuum c}
\end{equation}
\end{subequations}
defines a real antisymmetric bilinear form 
specified by the real-valued $d\times d$ antisymmetric matrix
$M^{(\bs{\ast})}$. When $d$ is even, we assume that
$M^{(\bs{\ast})}$ is invertible. When $d$ is odd,
$M^{(\bs{\ast})}$ has at least one vanishing eigenvalue
and is thus not invertible. Observe that
\begin{equation}
\hat{\rho}(\bs{q})=
\hat{\varrho}(\bs{q};-\bs{q}).
\label{eq: relating rho and varrho continuum}
\end{equation}

We are going to prove that (1) the family
$\hat{\varrho}(\bs{q};\bs{G})$
labeled by the pair $\bs{q}$ and $\bs{G}$ from $\mathbb{R}^{d}$
obeys the $U(1)$ algebra~(\ref{eq: def magnetic algebra}), and
(2) in even-dimensional space,
the family
$\hat{\varrho}(\bs{q};\bs{G})$
labeled by the pair $\bs{q}$ and $\bs{G}$ from $\mathbb{R}^{d}$
is complete.

\begin{widetext}

\textit{Proof of closure.}
We define
\begin{subequations}
\label{eq: algebra rho g p,q continuum}
\begin{equation}
\Gamma(\bs{q},\bs{q}',\bs{p};\bs{G},\bs{G}'):=
\Phi(\bs{q},\bs{q}'+\bs{p};\bs{G})
+
\Phi(\bs{q}',\bs{p};\bs{G}')
-
\Phi(\bs{q}+\bs{q}',\bs{p};\bs{G}+\bs{G}')
\label{eq: algebra rho g p,q continuum a}
\end{equation}
in terms of which Eq.%
~(\ref{eq: commutator two rho f's continuum})
can be rewritten as
\begin{equation}
\begin{split}
\left[
\hat{\varrho}(\bs{q} ;\bs{G} ),
\hat{\varrho}(\bs{q}';\bs{G}')
\right]=&\,
\int\limits_{\bs{p}}
\left[
e^{
\mathrm{i}\,
\Gamma(\bs{q},\bs{q}',\bs{p};\bs{G},\bs{G}')
  }
-
(
\bs{q}\leftrightarrow\bs{q}'
\hbox{ and }
\bs{G}\leftrightarrow\bs{G}'
)
\right]\,
e^{
+\mathrm{i}\,\Phi(\bs{q}+\bs{q}',\bs{p};\bs{G}+\bs{G}')
  }\,
\hat{c}^{\dag}(\bs{q}+\bs{q}'+\bs{p})\,
\hat{c}(\bs{p}).
\end{split}
\label{eq: algebra rho g p,q continuum b}
\end{equation}
\end{subequations}

Since 
\begin{subequations}
\label{eq: evaluation Upsilon continuum}
\begin{equation}
\begin{split}
\Gamma(\bs{q},\bs{q}',\bs{p};\bs{G},\bs{G}')=&\,
\left(\bs{q}+\frac{1}{2}\bs{G}\right)\,
\bs{\ast}\,
\left(\bs{q}'+\frac{1}{2}\bs{G}'\right)
-
\frac{1}{4}
\bs{G}\,\bs{\ast}\,\bs{G}'
\equiv
\Upsilon(\bs{q},\bs{q}';\bs{G},\bs{G}')
\end{split}
\end{equation}
is independent of $\bs{p}$ and antisymmetric under
$\bs{q}\leftrightarrow\bs{q}'$
and
$\bs{G}\leftrightarrow\bs{G}'$,
\begin{equation}
\Upsilon(\bs{q},\bs{q}';\bs{G},\bs{G}')=
-
\Upsilon(\bs{q}',\bs{q};\bs{G}',\bs{G}),
\end{equation}
\end{subequations}
the algebra~(\ref{eq: algebra rho g p,q continuum b})
closes to
\begin{subequations}
\label{eq: closure of algebra in continuum}
\begin{equation}
\left[
\hat{\varrho}(\bs{q} ;\bs{G} ),
\hat{\varrho}(\bs{q}';\bs{G}')
\right]=
F(\bs{q},\bs{q}';\bs{G},\bs{G}')\,
\hat{\varrho}(\bs{q}+\bs{q}';\bs{G}+\bs{G}')
\end{equation}
with the structure constant
\begin{equation}
\begin{split}
F(\bs{q},\bs{q}';\bs{G},\bs{G}')=&\,
e^{
\mathrm{i}\,
\Upsilon(\bs{q},\bs{q}';\bs{G},\bs{G}')
  }
-
(
\bs{q}\leftrightarrow\bs{q}'
\hbox{ and }
\bs{G}\leftrightarrow\bs{G}'
)
=
2\mathrm{i}\,
\sin\,\Upsilon(\bs{q},\bs{q}';\bs{G},\bs{G}').
\end{split}
\end{equation}
\end{subequations}

\textit{Proof of completeness.}
Choose any function
$f:\mathbb{R}^{d}\times\mathbb{R}^{d}\longrightarrow\mathbb{C}$
such that the Fourier transform
\begin{equation}
f(\bs{q},\bs{p})=:
\bar{f}(\bs{q},\bs{p})\,
e^{
\mathrm{i}\,\bs{q}\,\bs{\ast}\,\bs{p}
  }
=:
\left(
\int\limits_{\bs{G}}\,
e^{
+\mathrm{i}\,
\bs{G}\,\bs{\ast}\,\bs{p}
  }
\,
\tilde{f}(\bs{q},\bs{G})
\right)
e^{
\mathrm{i}\,\bs{q}\,\bs{\ast}\,\bs{p}
  }
\end{equation}
is well defined. 
For the second equality to be true for arbitrary functions
$\bar{f}(\bs{q},\cdot):\mathbb{R}^{d}\to\mathbb{C}$
with the well-defined Fourier transform
$\tilde{f}(\bs{q},\cdot):\mathbb{R}^{d}\to\mathbb{C}$,
the square matrix $M^{(\bs{\ast})}$
that defines the $\bs{\ast}$ product
must be invertible and thus have an even number $d$ of rows (columns).
Indeed, the rank of an antisymmetric matrix
$M^{(\bs{\ast})}$ is necessarily even. Hence, 
in odd-dimensional space,  $M^{(\bs{\ast})}$ is never
invertible as it has at least one vanishing eigenvalue.
This means that the $\bs{\ast}$ Fourier transform
$
\int\limits_{\bs{G}}\,
e^{
+\mathrm{i}\,
\bs{G}\,\bs{\ast}\,\bs{p}
  }
\,
\tilde{h}(\bs{G})
$
is at best a function of $d-1$ coordinates of $\bs{p}$
if $d$ is odd. For completeness to hold, it is thus necessary
that $d$ be even, which we now assume. A sufficient condition
for completeness to hold is that the linear space spanned
by the operators~(\ref{eq: def T's continuum})
is limited to the coherent superpositions of the
form~(\ref{eq: def rho f continuum}) 
such that the function
$\bar{f}(\bs{q},\cdot):\mathbb{R}^{d}\to\mathbb{C}$ has a Fourier
transform for any given momentum $\bs{q}$. With the help of
Eq.~(\ref{eq: def rgo G q continuum b}),
we can then write
\begin{equation}
f(\bs{q},\bs{p})=
\int\limits_{\bs{G}}\,
\tilde{f}(\bs{q},\bs{G})\,
e^{
\mathrm{i}\,\bs{q}\,\bs{\ast}\,\bs{G}/2
  }\,
e^{
\mathrm{i}\Phi(\bs{q},\bs{p};\bs{G})
  }.
\end{equation}
In turn and with the help of
Eq.~(\ref{eq: def rho f continuum}),
we conclude with
\begin{equation}
\hat{\varrho}^{f}(\bs{q})=
\int\limits_{\bs{p}}\,
f(\bs{q},\bs{p})\,
\hat{c}^{\dag}(\bs{q}+\bs{p})\,
\hat{c}^{\,}(\bs{p})
=
\int\limits_{\bs{G}}\,
\tilde{f}(\bs{q},\bs{G})
e^{
\mathrm{i}\,\bs{q}\,\bs{\ast}\,\bs{G}/2
  }
\int\limits_{\bs{p}}\,
e^{
\mathrm{i}\Phi(\bs{q},\bs{p};\bs{G})
  }\,
\hat{c}^{\dag}(\bs{q}+\bs{p})\,
\hat{c}^{\,}(\bs{p})
=
\int\limits_{\bs{G}}\,
\tilde{f}(\bs{q},\bs{G})\,
e^{
\mathrm{i}\,\bs{q}\,\bs{\ast}\,\bs{G}/2
  }\,
\hat{\varrho}(\bs{q};\bs{G}).
\end{equation}

\end{widetext}

\section{
The case of the lattice
        }
\label{sec: lattice}

We begin with some notation.
Let $\Lambda$ be a Bravais lattice 
and $\Lambda^{\star}$ be its dual.
Sites in $\Lambda$ are denoted by $\bs{r}$, and sites in $\Lambda^{\star}$
are denoted
by $\bs{G}$. The first Brillouin zone is denoted $\Omega^{\,}_{\mathrm{BZ}}$;
it contains the origin of $\mathbb{R}^{d}$. We shall decompose
$\mathbb{R}^{d}$ into a set of shifted Brillouin zones
$\Omega^{\bs{G}}_{\mathrm{BZ}}$
obtained by translation of
$\Omega^{\,}_{\mathrm{BZ}}$ by
$\bs{G}\,\in\Lambda^{\star}$,
\begin{equation}
\mathbb{R}^{d}=
\bigcup_{\bs{G}\,\in\Lambda^{\star}}
\Omega^{\bs{G}}_{\mathrm{BZ}}.
\label{eq: union all BZs}
\end{equation}
Sites in $\Omega^{\,}_{\mathrm{BZ}}$ are denoted $\bs{k}$, $\bs{q}$,
and $\bs{p}$. If $\bs{q}$ and $\bs{p}$ belong to the Brillouin zone
$\Omega^{\,}_{\mathrm{BZ}}$, this might not be the case for
$\bs{q}+\bs{p}$. There is a unique
$\bs{G}^{\,}_{\bs{q}+\bs{p}}\,\in\Lambda^{\star}$ such that 
$\bs{q}+\bs{p}\,\in\Omega^{\bs{G}^{\,}_{\bs{q}+\bs{p}}}_{\mathrm{BZ}}$.
Correspondingly,
$\bs{q}+\bs{p}-\bs{G}^{\,}_{\bs{q}+\bs{p}}\,\in\Omega^{\,}_{\mathrm{BZ}}$.
We shall use the notation
\begin{subequations}
\begin{equation}
[\bs{q}+\bs{p}]^{\,}_{\mathrm{BZ}}\equiv
\bs{q}+\bs{p}-\bs{G}^{\,}_{\bs{q}+\bs{p}}
\,\in\Omega^{\,}_{\mathrm{BZ}}.
\label{eq: def noation [...]BZ}
\end{equation}
Two observations are pertinent to what follows.
First, the bracket~(\ref{eq: def noation [...]BZ})
obeys the nesting rule
\begin{equation}
\left[\vphantom{\Big[}
\bs{q}'+[\bs{q}+\bs{p}]^{\,}_{\mathrm{BZ}}
\right]^{\,}_{\mathrm{BZ}}=
[
\bs{q}
+
\bs{q}'
+
\bs{p}
]^{\,}_{\mathrm{BZ}}
\label{eq: nested bracket}
\end{equation}
\end{subequations}
for any triplet 
$\bs{q}$,
$\bs{q}'$,
and
$\bs{p}$
from the first Brillouin zone.
Second,
if we hold 
$\bs{q}\,\in\Omega^{\ }_{\mathrm{BZ}}$ 
fixed and vary
$\bs{p}$
across the Brillouin zone $\Omega^{\ }_{\mathrm{BZ}}$,
the unique reciprocal wave vector 
$\bs{G}^{\,}_{\bs{q}+\bs{p}}\,\in\Lambda^{\star}$
such that 
$\bs{q}+\bs{p}-\bs{G}^{\,}_{\bs{q}+\bs{p}}\,\in\Omega^{\,}_{\mathrm{BZ}}$
defines an implicit function of $\bs{q}$ that is piecewise constant 
with discontinuous jumps each time $\bs{q}+\bs{p}$ crosses
the boundary separating neighboring Brillouin zones.

We define the fermionic Fock space $\mathfrak{F}$ with the help
of the algebra
\begin{equation}
\begin{split}
&
\left\{
\hat{c}^{\,  }_{\bs{k} +\bs{G} },
\hat{c}^{\dag}_{\bs{k}'+\bs{G}'}
\right\}=
\delta^{\ }_{\bs{k},\bs{k}'},
\\
&
\left\{
\hat{c}^{\,}_{\bs{k} +\bs{G} },
\hat{c}^{\,}_{\bs{k}'+\bs{G}'}
\right\}=
\left\{
\hat{c}^{\dag}_{\bs{k} +\bs{G} },
\hat{c}^{\dag}_{\bs{k}'+\bs{G}'}
\right\}=0
\end{split}
\label{eq: def fermion algebra lattice}
\end{equation}
for any pair $\bs{k}$ and $\bs{k}'$ from the Brillouin zone 
$\Omega^{\,}_{\mathrm{BZ}}$
and any pair  $\bs{G}$ and $\bs{G}'$ from the dual lattice
$\Lambda^{\star}$.

The linear space of fermionic bilinears that we
study is spanned by the basis
\begin{subequations}
\begin{equation}
\hat{T}^{\,}_{\bs{q}^{\,}_{1},\bs{q}^{\,}_{2}}:=
\hat{c}^{\dag}_{\bs{q}^{\,}_{1}}\,
\hat{c}^{\,  }_{\bs{q}^{\,}_{2}},
\label{eq: ded T12 family}
\end{equation}
which obeys the algebra
\begin{equation}
\left[
\hat{T}^{\,}_{\bs{q}^{\,}_{1},\bs{q}^{\,}_{2}},
\hat{T}^{\,}_{\bs{q}^{\prime}_{1},\bs{q}^{\prime}_{2}}
\right]=
\delta^{\,}_{\bs{q}^{\,}_{2},\bs{q}^{\prime}_{1}}\,
\hat{T}^{\,}_{\bs{q}^{\,}_{1},\bs{q}^{\prime}_{2}}
-
\delta^{\,}_{\bs{q}^{\,}_{1},\bs{q}^{\prime}_{2}}\,
\hat{T}^{\,}_{\bs{q}^{\prime}_{1},\bs{q}^{\,}_{2}}
\label{eq: algebra T's}
\end{equation}
\end{subequations}
for any quadruple 
$\bs{q}^{\,}_{1}$,
$\bs{q}^{\,}_{2}$,
$\bs{q}^{\prime}_{1}$,
and
$\bs{q}^{\prime}_{2}$
from the Brillouin zone.

For any $\bs{q}$ from the  Brillouin zone 
$\Omega^{\,}_{\mathrm{BZ}}$
and for any function
$
f:\Omega^{\,}_{\mathrm{BZ}}\times\Omega^{\,}_{\mathrm{BZ}}
\longrightarrow\mathbb{C}
$,
define
\begin{equation}
\hat{\varrho}^{f}_{\bs{q}}:=
\sum_{\bs{p}\,\in\Omega^{\,}_{\mathrm{BZ}}}
f^{\,}_{\bs{q},\bs{p}}\,
\hat{c}^{\dag}_{[\bs{q}+\bs{p}]^{\,}_{\mathrm{BZ}}}\,
\hat{c}^{\,  }_{\bs{p}}.
\label{eq: def varrho lattice}
\end{equation}
There follows the algebra
[with the help of Eq.~(\ref{eq: nested bracket})]
\begin{equation}
\begin{split}
\left[
\hat{\varrho}^{f }_{\bs{q} },
\hat{\varrho}^{f'}_{\bs{q}'}
\right]=&\,
\sum_{\bs{p}\,\in\Omega^{\,}_{\mathrm{BZ}}}
\left[
f^{\,    }_{
\bs{q},[\bs{q}'+\bs{p}]^{\,}_{\mathrm{BZ}}
           }\,
f^{\prime}_{
\bs{q}',\bs{p}
           }\,
-
(
\bs{q},f\leftrightarrow\bs{q}',f'
)
\right]
\\
&\,
\times
\hat{c}^{\dag}_{
[\bs{q}+\bs{q}'+\bs{p}]^{\,}_{\mathrm{BZ}}
               }\,
\hat{c}^{\,  }_{\bs{p} }
\end{split}
\label{eq: commutator two rhos}
\end{equation}
for any pair of momenta $\bs{q}$ and $\bs{q}'$
from the Brillouin zone
and for any pair of functions $f$ and $f'$.

The choice $f^{\,}_{\bs{q},\bs{p}}=1$
for any pair $\bs{q}$ and $\bs{p}$
from the Brillouin zone
defines the momentum representation of the 
local density operator
\begin{subequations}
\begin{equation}
\hat{\rho}^{\,}_{\bs{q}}:=
\sum_{\bs{p}\in\Omega^{\,}_{\mathrm{BZ}}}
\hat{c}^{\dag}_{[\bs{q}+\bs{p}]^{\,}_{\mathrm{BZ}}}\,
\hat{c}^{\,}_{\bs{p}}.
\label{eq: def rho lattice}
\end{equation}
Any pair thereof commutes as
\begin{equation}
\left[
\hat{\rho}^{\,}_{\bs{q} },
\hat{\rho}^{\,}_{\bs{q}'}
\right]=0.
\label{eq: algebra densities lattice}
\end{equation}
\end{subequations}

Another choice of the function $f$ is made with
the family 
\begin{subequations}
\label{eq: def Phi G q p}
\begin{equation}
\hat{\varrho}^{\bs{G} }_{\bs{q}}:=
\sum_{\bs{p}\,\in\Omega^{\,}_{\mathrm{BZ}}}
e^{+\mathrm{i}\,\Phi^{\bs{G}}_{\bs{q},\bs{p}}}\,
\hat{c}^{\dag}_{[\bs{q}+\bs{p}]^{\,}_{\mathrm{BZ}}}\,
\hat{c}^{\,  }_{\bs{p}}
\label{eq: def varrho G q}
\end{equation}
for any $\bs{G}$ from the dual lattice and $\bs{q}$
from the Brillouin zone, where 
\begin{equation}
\Phi^{\bs{G}}_{\bs{q},\bs{p}}:=
\frac{1}{2\pi}
\left[
(\bs{q}+\bs{G})\,\bs{\ast}\,\bs{p}
-
(\bs{q}+\bs{p}+\bs{G})
\,\bs{\ast}\,
\bs{G}^{\,}_{\bs{q}+\bs{p}+\bs{G}}
\right]
\label{eq: def Phi G q p a}
\end{equation}
\end{subequations}
and the $d\times d$ matrix $M^{(\bs{\ast})}_{\Lambda}$ 
that defines the $\bs{\ast}$ product is 
antisymmetric, as was the case in the continuum, 
but with the restriction that
\begin{equation}
\frac{1}{2\pi}\,\bs{G}\,\bs{\ast}\,\bs{G}'=
0 \hbox{ mod } 2\pi,
\qquad
\forall\bs{G},\bs{G}'\in\Lambda^{\star},
\label{eq: * product set by Bravais lattice}
\end{equation}
to accommodate the $d$-dimensional Bravais lattice $\Lambda$.
When $d$ is even, $M^{(\bs{\ast})}_{\Lambda}$ 
has a nonvanishing determinant by assumption.
As announced, the $U(1)$ algebra
\begin{equation}
\left[
\hat{\varrho}^{\bs{G} }_{\bs{q} },
\hat{\varrho}^{\bs{G}'}_{\bs{q}'}
\right]=
2\mathrm{i}\,
\sin\,
\left(
\frac{
(\bs{q}+\bs{G})\,\bs{\ast}\,(\bs{q}'+\bs{G}')
     }
     {
2\pi
     }
\right)\,
\hat{\varrho}^{\bs{G} +\bs{G}'}_{\bs{q}+\bs{q}'}
\label{eq: closed GMP algebra on the lattice}
\end{equation}
follows for any quadruple 
$\bs{q}$,
$\bs{q}'$,
$\bs{G}$,
and
$\bs{G}^{\prime}$.
The proof of 
Eq.~(\ref{eq: closed GMP algebra on the lattice})
is technically more involved than that of 
Eq.~(\ref{eq: closure of algebra in continuum})
as one needs to account for the restriction on momenta 
to the first Brillouin zone. 
For this reason, we refer the reader to Appendix%
~\ref{appsec: Proof closure algebra}
for the details of the proof.

To prove completeness,
we assume that the dimensionality $d$ is even
for the same reasons as given below Eq.%
~(\ref{eq: def rgo G q continuum c}).
One verifies that
\begin{subequations}
\label{eq: evaluation Phi q and p lattice}
\begin{equation}
\Phi^{\bs{G}}_{\bs{q},\bs{p}}=
\Theta^{\,}_{\bs{q},\bs{p}}
+
\frac{
\bs{G}\,\bs{\ast}\,\bs{p}
-
\bs{p}\,\bs{\ast}\,\bs{G}
     }
     {
2\pi
     }
+
\Theta^{\bs{G}}_{\bs{q}}
+
\hbox{ mod }
2\pi,
\end{equation}
where the function
\begin{equation}
2\pi\,
\Theta^{\,}_{\bs{q},\bs{p}}:=
\bs{q}\,\bs{\ast}\,\bs{p}
-
(\bs{q}+\bs{p})\,\bs{\ast}\,\bs{G}^{\,}_{\bs{q}+\bs{p}}
\label{eq: evaluation Phi q and p lattice a}
\end{equation}
is independent of $\bs{G}$, while the function
\begin{equation}
2\pi\,
\Theta^{\bs{G}}_{\bs{q}}:=
-
\bs{q}\,\bs{\ast}\,\bs{G}
\label{eq: evaluation Phi q and p lattice b}
\end{equation}
\end{subequations}
is independent of $\bs{p}$.
We will use the fact that
\begin{equation}
\Theta^{\,}_{\bs{q}=\bs{0},\bs{p}}=
\Theta^{\bs{G}}_{\bs{q}=\bs{0}}=0
\label{eq: Theta-q-p and Theta-q-G can vanish if q=0}
\end{equation}
in Sec.~\ref{subsec: Normal ordering and the bare band width}
and Appendix~\ref{appendix: Equation MAIN RESULT}.
We define the function 
$
\bar{f}:\Omega^{\,}_{\mathrm{BZ}}\times\Omega^{\,}_{\mathrm{BZ}}
\longrightarrow\mathbb{C}
$
by
\begin{equation}
f^{\,}_{\bs{q},\bs{p}}=:
\bar{f}^{\,}_{\bs{q},\bs{p}}\,
e^{+\mathrm{i}\,\Theta^{\,}_{\bs{q},\bs{p}}}.
\label{eq: from f to tilde f lattice}
\end{equation}
We then use the Fourier expansion
\begin{equation}
\bar{f}^{\,}_{\bs{q},\bs{p}}=:
\sum_{\bs{G}\,\in\Lambda^{\star}}
\tilde{f}^{\bs{G}}_{\bs{q}}\,
e^{
+\mathrm{i}\,
\left(
\bs{G}\,\bs{\ast}\,\bs{p}
-
\bs{p}\,\bs{\ast}\,\bs{G}
\right)
/(2\pi)
  }
\label{eq: Fourier trsf lattice}
\end{equation}
to do the following manipulations:
\begin{eqnarray}
f^{\,}_{\bs{q},\bs{p}}&=&
e^{+\mathrm{i}\,\Theta^{\,}_{\bs{q},\bs{p}}}\,
\bar{f}^{\,}_{\bs{q},\bs{p}}
\nonumber\\
&=&
e^{+\mathrm{i}\,\Theta^{\,}_{\bs{q},\bs{p}}}\,
\left[
\sum_{\bs{G}\,\in\Lambda^{\star}}
\tilde{f}^{\bs{G}}_{\bs{q}}
e^{
+\mathrm{i}\,
\left(
\bs{G}\,\bs{\ast}\,\bs{p}
-
\bs{p}\,\bs{\ast}\,\bs{G}
\right)
/(2\pi)
  }
\right]
\nonumber\\
&=&
\sum_{\bs{G}\,\in\Lambda^{\star}}
\tilde{f}^{\bs{G}}_{\bs{q}}
e^{
+
\mathrm{i}\,
\Theta^{\,}_{\bs{q},\bs{p}}
+
\mathrm{i}\,
\left(
\bs{G}\,\bs{\ast}\,\bs{p}
-
\bs{p}\,\bs{\ast}\,\bs{G}
\right)
/(2\pi)
+
\mathrm{i}\,
\Theta^{\bs{G}}_{\bs{q}}
-
\mathrm{i}\,
\Theta^{\bs{G}}_{\bs{q}}
  }
\nonumber\\
&=&
\sum_{\bs{G}\,\in\Lambda^{\star}}
\underbrace{
\tilde{f}^{\bs{G}}_{\bs{q}}
e^{
-
\mathrm{i}\,
\Theta^{\bs{G}}_{\bs{q}}
  }
           }_{\hbox{independent of $\bs{p}$}}
\times\,
e^{
+
\mathrm{i}\,
\Phi^{\bs{G}}_{\bs{q},\bs{p}}
  }.
\label{eq: correct fourier expansion f}
\end{eqnarray}
Inserting Eq.~(\ref{eq: correct fourier expansion f})
into Eq.~(\ref{eq: def varrho lattice})
gives
\begin{eqnarray}
\hat{\varrho}^{f}_{\bs{q}}&=&
\sum_{\bs{G}\,\in\Lambda^{\star}}
\tilde{f}^{\bs{G}}_{\bs{q}}
e^{
-
\mathrm{i}\,
\Theta^{\bs{G}}_{\bs{q}}
  }
\,
\left(
\sum_{\bs{p}\,\in\Omega^{\,}_{\mathrm{BZ}}}
e^{
+
\mathrm{i}\,
\Phi^{\bs{G}}_{\bs{q},\bs{p}}
  }
\,
\hat{c}^{\dag}_{[\bs{q}+\bs{p}]^{\,}_{\mathrm{BZ}}}\,
\hat{c}^{\,  }_{\bs{p}}
\right)
\nonumber\\
&=&
\sum_{\bs{G}\,\in\Lambda^{\star}}
\tilde{f}^{\bs{G}}_{\bs{q}}
e^{
-
\mathrm{i}\,
\Theta^{\bs{G}}_{\bs{q}}
  }\,
\hat{\varrho}^{\bs{G}}_{\bs{q}},
\label{eq: eq stating completeness}
\end{eqnarray}
where we made use of definition~(\ref{eq: def varrho G q})
to reach the last equality. Completeness has thus been proved if
the space of functions $f$ is restricted to those for which
Fourier transform~(\ref{eq: Fourier trsf lattice}) exists.

\section{
Discussion
        }
\label{sec: Discussion}

As pointed out by Murthy and Shankar,
the magnetic translation algebra is not limited
to situations in which time-reversal symmetry is broken.
From the point of view of many-body physics,
the generators of the magnetic translation algebra
can also be thought of as special coherent superpositions
of particle-hole excitations.
As such they are always present in the 
many-body Fock space. 

If time-reversal symmetry is 
either explicitly or spontaneously broken, 
it is plausible that these excitations might
be selected by the many-body interactions to play
an important role at low energies and long distances.
However, the breaking of time-reversal symmetry alone
is no guarantee for the FQHE. The selection of a
ground state supporting the FQHE is a subtle compromise
between the kinetic energy and the interactions.

If time-reversal symmetry is 
neither explicitly nor spontaneously broken,
it is harder to imagine that these excitations
are of relevance to the low-energy and long-distance
properties of interacting electrons.

With this motivation in mind, we are going to discuss 
the following two cases.

\textit{(a) The $f$-sum rule.}
We begin in Sec.~\ref{subsec: f sum rule}
with the case of interacting electrons
in the continuum limit without explicit
breaking of time-reversal symmetry and 
for which spontaneous symmetry breaking
of time-reversal symmetry is not anticipated. 
This situation is the one expected 
if electrons interact through sufficiently weak density-density 
interactions. We are going to show how to recover the $f$-sum rule
when we choose to represent the many-body Hamiltonian
in terms of the generators~(\ref{eq: def rgo G q continuum})
of the magnetic translation algebra
for any even dimension $d$ of space.

This exercise serves two purposes.
First, it gives us the confidence that we can solve an interacting
problem devoid of any magnetic field using 
the magnetic translation algebra, 
i.e., using a technology that is geared to the presence of
a magnetic field. We find this result remarkable.
Second, it is a warning against blindly performing a mean-field approximation
of the Hamiltonian, when represented in terms of the generators%
~(\ref{eq: def rgo G q continuum}),
that delivers the FQHE. In other words, one should be cautious when
using the magnetic translation algebra in an approximate fashion
to predict a FQHE, for such treatments can predict a FQHE when
none is known to occur.

\textit{(b) FCIs at intermediate rather than strong couplings.}
To illustrate the delicate competition between the
kinetic energy and the interactions, we consider in Sec.%
~\ref{subsec: Normal ordering and the bare band width}
a Hamiltonian describing a band insulator 
to which we add density-density interactions that preserve translation
invariance. We represent the projection of this Hamiltonian
onto a single band in terms of the generators~(\ref{eq: def Phi G q p})
for any even dimension $d$ of the Bravais lattice.
In doing so, we are going to show
that normal ordering can change the bare bandwidth
by a value comparable to the characteristic energy for the 
interactions. Hence, if the bare bandwidth is
smaller than the characteristic energy for the interactions,
as is usually believed to be necessary to stabilize a FCI,
normal ordering can be an effect of order 1.

As an application of this result, we 
consider any projected and normal-ordered
Hamiltonian $\hat{H}$ describing itinerant
fermions in a flat band carrying a nonvanishing Chern number
and interacting though a density-density interaction
that preserves translation invariance.
We assume that $\hat{H}$ supports
a FCI as the ground state at the partial filling 
$0<\nu<1$ of the flat band.
A particle-hole transformation turns the normal-ordered
$\hat{H}$ into $\widetilde{H}$,
whereby $\widetilde{H}$ must support 
a FCI made of holes as the ground state at the partial filling 
$\widetilde{\nu}=1-\nu$. What is remarkable is that
the projected Hamiltonian $\widetilde{H}$, when decomposed into
a one-body term and a normal-ordered interaction, can be thought
of as describing holes with a genuine dispersion and interacting
through a normal-ordered density-density interaction sharing
the same functional form as $\hat{H}$. The dispersion 
of the holes is genuine
because its width is generically nonvanishing and of the order of
the characteristic interaction strength times a numerical factor
of geometrical origin. Indeed, this numerical factor arises because
of the geometry induced by the overlaps between pairs
of Bloch states from the original flat band.%
~\cite{Page87}
When these overlaps
are constant, as is the case in the FQHE, this numerical factor
vanishes so that $\widetilde{H}$ can also be assigned a flat band.
When these overlaps are functions of both the relative and 
center-of-mass momenta of the pair of Bloch states, then 
this numerical factor can be nonvanishing.

That this numerical factor can be of order unity, and thus
matters in a crucial way in order to stabilize the FCI at 
the filling fraction $\widetilde{\nu}$,
can be inferred from the following numerical results. 

In Ref.~\onlinecite{Neupert11a}, a band insulator
with two flat bands supporting the Chern numbers $\pm1$
was shown to support a FCI phase at the filling fraction
$1/3$ in the presence of a repulsive nearest-neighbor
density-density interaction projected onto the lower flat band.
In Ref.~\onlinecite{Regnault11a}, 
the \textit{same} band insulator
was shown to support the \textit{same} FCI phase 
at the \textit{same} filling fraction
$1/3$ in the presence of a \textit{different}
interaction, namely, the 
repulsive nearest-neighbor density-density interaction
projected onto the lower flat band and then \textit{normal ordered}.
Hence, at the filling fraction $1/3$, the FCI phase
is robust to whether the projected interaction is normal ordered or not.
In Ref.~\onlinecite{Neupert11b}, the same model as in
Ref.~\onlinecite{Neupert11a} was also shown to support
a FCI phase at the filling fraction $2/3$.
However, no evidence for a topological phase was
found at the filling fraction $2/3$ using the \textit{normal ordered}
projected interaction in Ref.~\onlinecite{Regnault11a}. 
Hence, at the filling fraction $2/3$, the FCI phase
either is not selected as the ground state or is very close 
to a phase transition to a phase without topological order
when the projected interaction is normal ordered, 
while the FCI phase is selected as the ground state when 
the projected interaction is not normal ordered. We conclude
that the characteristic bandwidth of the
one-body term that is generated by normal ordering the repulsive
nearest-neighbor density-density interaction
must be of the same order as the characteristic energy 
scale of the interaction.

Both quantitative examples are 
consistent with the fact that interactions projected onto a 
single Chern band can induce one-body terms that can significantly 
alter the bandwidth of lattice Hamiltonians for itinerant fermions.

\subsection{
$f$-sum rule
           }
\label{subsec: f sum rule}

The $f$-sum rule holds for electrons with mass $m$
and the quadratic dispersion 
\begin{subequations}
\label{eq: f sum rule}
\begin{equation}
\varepsilon(\bs{p})\,:=
\frac{\bs{p}^{2}}{2m}
\label{eq: f sum rule a}
\end{equation}
subjected to any one-body potential $V$ and
interacting with any translation-invariant
density-density interaction $U$
in any dimension $d$. 
Pines and Nozi\`eres presented a derivation thereof
in Ref.~\onlinecite{Pines66}
that hinges on the fact that the operator identity
\begin{equation}
\Big[
\hat{\rho}(\bs{q}),
[\hat{H},\hat{\rho}(\bs{-q})]
\Big]=
\frac{\bs{q}^{2}}{m}\,
\hat{N}
\label{eq: f sum rule b}
\end{equation}
holds for any momentum $\bs{q}\in\mathbb{R}^{d}$.
Here,
\begin{equation}
\hat{N}:=
\int\limits_{\bs{p}}\,
\hat{c}^{\dag}(\bs{p})\,
\hat{c}(\bs{p})
\label{eq: f sum rule c}
\end{equation}
is the conserved particle number operator,
and the many-body Hamiltonian 
$
\hat{H}:=
\hat{H}^{\,}_{0}
+
\hat{H}^{\,}_{V}
+
\hat{H}^{\,}_{U}
$
is the sum of the dispersion
\begin{equation}
\hat{H}^{\,}_{0}:=
\int\limits_{\bs{p}}
\varepsilon(\bs{p})\,
\hat{c}^{\dag}(\bs{p})\,
\hat{c}(\bs{p}),
\label{eq: f sum rule d}
\end{equation}
the one-body potential
\begin{equation}
\hat{H}^{\,}_{V}:=
\int\limits_{\bs{q}}
V(\bs{q})\,
\hat{\rho}(-\bs{q}),
\label{eq: f sum rule e}
\end{equation}
and the two-body potential
\begin{equation}
\hat{H}^{\,}_{U}:=
\int\limits_{\bs{q}} 
U(\bs{q})\;
\hat{\rho}(\bs{q})\,
\hat{\rho}(\bs{-q}).
\label{eq: f sum rule f}
\end{equation}
\end{subequations}
The only nonvanishing
contribution to the nested commutator in Eq.%
~(\ref{eq: f sum rule b})
arises from the quadratic dispersion $\hat{H}^{\,}_{0}$
in view of the Abelian algebra%
~(\ref{eq: algebra densities continuum}).
Equation~(\ref{eq: f sum rule b})
follows from the algebra~(\ref{eq: def fermion algebra continuum}).

As a sanity check, we are going to verify
Eq.%
~(\ref{eq: f sum rule b})
for any even dimension $d$
with the help of the magnetic translation algebra
\begin{equation}
\begin{split}
\left[
\hat{\varrho}(\bs{q} ;\bs{G} ),
\hat{\varrho}(\bs{q}';\bs{G}')
\right]=&\,
2\mathrm{i}\,
\sin\Upsilon(\bs{q},\bs{q}';\bs{G},\bs{G}')
\\
&\,
\times\,
\hat{\varrho}(\bs{q}+\bs{q}';\bs{G}+\bs{G}'),
\end{split}
\label{eq: magnetic algebra for f sum rule}
\end{equation}
where $\Upsilon(\bs{q},\bs{q}';\bs{G},\bs{G}')$
is defined in Eq.~(\ref{eq: evaluation Upsilon continuum}).
We shall only evaluate the contribution from
the quadratic dispersion~(\ref{eq: f sum rule d}).

First, we recall that
$\hat{\rho}(\bs{q})=\hat{\varrho}(\bs{q};\bs{-q})$ 
according to
Eq.~(\ref{eq: relating rho and varrho continuum}).
Second, we expand $\hat{H}^{\,}_{0}$
in terms of the magnetic translation densities
$\hat{\varrho}(\bs{q};\bs{G})$,
\begin{subequations}
\label{eq: rep H0 varrho}
\begin{eqnarray}
\hat{H}^{\,}_{0}&=&
\int\limits_{\bs{G}} 
\tilde{\varepsilon}(\bs{G})\;
\hat{\varrho}(\bs{q}=0;\bs{G}),
\label{eq: rep H0 varrho a}
\end{eqnarray}
where
\begin{eqnarray}
\tilde{\varepsilon}(\bs{G})&=&
\int\limits_{\bs{p}} 
e^{
-\mathrm{i}\,\bs{G}\bs{\ast}\bs{p}
  }\,
\varepsilon(\bs{p}).
\label{eq: rep H0 varrho b}
\end{eqnarray}
\end{subequations}
It is with Eq.~(\ref{eq: rep H0 varrho b})
that we made use of $d$ being even.
\begin{widetext}

\noindent
Third, we make a first use of 
Eq.~(\ref{eq: magnetic algebra for f sum rule})
to evaluate the internal commutator
\begin{eqnarray}
\Big[\hat{\rho}(\bs{q}),[\hat H,\hat{\rho}(\bs{-q})]\Big]
&=&
\int\limits_{\bs{G}} 
\tilde{\varepsilon}(\bs{G})\;
\Big[
\hat{\varrho}(\bs{q};\bs{-q}),
[
\hat{\varrho}({0};\bs{G}),
\hat{\varrho}(\bs{-q};\bs{q})
]
\Big]
\nonumber\\
&=&
\int\limits_{\bs{G}} 
\tilde{\varepsilon}(\bs{G})\;
2\mathrm{i}\,
\sin\,\Upsilon(0,\bs{-q};\bs{G},\bs{q})\;
[
\hat{\varrho}(\bs{q};\bs{-q}),
\hat{\varrho}(\bs{-q};\bs{G+q})
].
\end{eqnarray}
We make a second use of 
Eq.~(\ref{eq: magnetic algebra for f sum rule})
to evaluate the external commutator
\begin{eqnarray}
\Big[\hat{\rho}(\bs{q}),[\hat H,\hat{\rho}(\bs{-q})]\Big]
&=&
\int\limits_{\bs{G}} 
\tilde{\varepsilon}(\bs{G})\;
2\mathrm{i}\,
\sin\,\Upsilon(0,\bs{-q};\bs{G},\bs{q})\;
2\mathrm{i}\,
\sin\,\Upsilon(\bs{q},\bs{-q};\bs{-q},\bs{G+q})\;
\hat{\varrho}(0;\bs{G})
\nonumber\\
&=&
\int\limits_{\bs{G}} 
\tilde{\varepsilon}(\bs{G})\;
\left[
2\mathrm{i}\,
\sin\,
\left(
\frac{\bs{q}\ast\bs{G}}{2}
\right)
\;
\right]^2
\;
\hat{\varrho}(0;\bs{G}).
\end{eqnarray}
The integral over $\bs{G}$ can now be performed,
\begin{eqnarray}
\Big[\hat{\rho}(\bs{q}),[\hat H,\hat{\rho}(\bs{-q})]\Big]
&=&
\int\limits_{\bs{G}} 
\tilde{\varepsilon}(\bs{G})\;
\left(
e^{+\mathrm{i}\,\bs{q}\ast\bs{G}}+e^{-\mathrm{i}\,\bs{q}\ast\bs{G}}
-2
\right)
\;
\int\limits_{\bs{p}} \;
e^{+\mathrm{i}\,\bs{G}\ast\bs{p}}\;
\hat{c}^{\dag}(\bs{p})\,
\hat{c}(\bs{p})
\nonumber\\
&=&
\int\limits_{\bs{p}} 
\frac{1}{2m}
\left(
|\bs{p+q}|^2
+
|\bs{p-q}|^2
-
2|\bs{p}|^2
\right)
\;
\hat{c}^{\dag}(\bs{p})\,
\hat{c}(\bs{p})
\nonumber\\
&=&
\frac{\bs{q}^{2}}{m}
\int\limits_{\bs{p}}\,
\hat{c}^{\dag}(\bs{p})\,
\hat{c}(\bs{p}).
\end{eqnarray}
Equation~(\ref{eq: f sum rule b})
follows from the definition~(\ref{eq: f sum rule c}).

\end{widetext}

\subsection{
Projected Hamiltonians and the importance of 
induced one-body terms
           }
\label{subsec: Normal ordering and the bare band width}

We begin with the generic lattice Hamiltonian
\begin{equation}
\hat{H}:=
\hat{H}^{\,}_{0}
+
\hat{H}^{\,}_{U},
\label{eq: generic lattice H}
\end{equation}
where the dimensionality $d$ of the lattice is 
assumed even.
Our goal is to understand how normal ordering 
of the interaction $\hat{H}^{\,}_{U}$ 
changes the bandwidth of the kinetic Hamiltonian
$\hat{H}^{\,}_{0}$. To this end, we need to choose
the representation in which we define
$\hat{H}^{\,}_{0}$
and
$\hat{H}^{\,}_{U}$. We will see that
the choice of the representation of
$\hat{H}$
can change the effects on
$\hat{H}^{\,}_{0}$
of normal ordering on
$\hat{H}^{\,}_{U}$.

The kinetic Hamiltonian is defined by
\begin{subequations}
\begin{equation}
\hat{H}^{\,}_{0}:=
\frac{1}{2}
\sum_{\bs{r},\bs{r}'\in\Lambda}
\sum_{\alpha,\alpha'}
\left(
\hat{\psi}^{\dag}_{\bs{r},\alpha}\,
t^{\alpha,\alpha'}_{\bs{r}-\bs{r}'}\,
\hat{\psi}^{\,}_{\bs{r}',\alpha'}
+
\mathrm{H.c.}
\right),
\end{equation}
where the hopping amplitudes
\begin{equation}
t^{\alpha,\alpha'}_{\bs{r}-\bs{r}'}=
\left(
t^{\alpha',\alpha}_{\bs{r}'-\bs{r}}
\right)^{*}
\end{equation}
\end{subequations}
decay exponentially fast with the separation between
any pair of sites $\bs{r}$ and $\bs{r}'$ from the lattice
$\Lambda$ and we have reinstated a finite number of
internal degrees of freedom
labeled by the \textit{orbital index} $\alpha$.
If $N$ denotes the number of sites in $\Lambda$,
we can perform the Fourier transformation to the band basis
in two steps. First, we do the Fourier transformation
\begin{subequations}
\begin{equation}
\hat{\psi}^{\dag}_{\bs{r},\alpha}=:
\sum_{\bs{p}\in\Omega^{\,}_{\mathrm{BZ}}}
\frac{
e^{
-\mathrm{i}\,\bs{p}\cdot\bs{r}
  }\,
     }
     {
\sqrt{N}
     }
\hat{\psi}^{\dag}_{\bs{p},\alpha},
\quad
\hat{\psi}^{\,}_{\bs{r},\alpha}=:
\sum_{\bs{p}\in\Omega^{\,}_{\mathrm{BZ}}}
\frac{
e^{
+\mathrm{i}\,\bs{p}\cdot\bs{r}
  }\,
     }
     {
\sqrt{N}
     }
\hat{\psi}^{\,}_{\bs{p},\alpha},
\end{equation}
in terms of which
\begin{equation}
\begin{split}
&
\hat{H}^{\,}_{0}=
\sum_{\bs{p}\in\Omega^{\,}_{\mathrm{BZ}}}
\sum_{\alpha,\alpha'}
\hat{\psi}^{\dag}_{\bs{p},\alpha}\,
\mathcal{H}^{\alpha,\alpha'}_{\bs{p}}\,
\hat{\psi}^{\,}_{\bs{p},\alpha'},
\\
&
\mathcal{H}^{\alpha,\alpha'}_{\bs{p}}:=
\sum_{\bs{r}\in\Lambda}
e^{-\mathrm{i}\bs{p}\cdot\bs{r}}\,
t^{\alpha,\alpha'}_{\bs{r}}.
\end{split}
\end{equation}
\end{subequations}
Second, for any given $\bs{p}$ from the Brillouin zone,
we do the unitary transformation
\begin{subequations}
\begin{equation}
\hat{\psi}^{\dag}_{\bs{p},\alpha}=:
\sum_{a}
\hat{c}^{\dag}_{\bs{p},a}
u^{\alpha\,*}_{\bs{p},a},
\qquad
\hat{\psi}^{\,}_{\bs{p},\alpha}=:
\sum_{a}
u^{\alpha}_{\bs{p},a}\,
\hat{c}^{\,}_{\bs{p},a},
\end{equation}
in terms of which
\begin{equation}
\hat{H}^{\,}_{0}=
\sum_{\bs{p}\in\Omega^{\,}_{\mathrm{BZ}}}
\sum_{a}
\hat{c}^{\dag}_{\bs{p},a}\,
\varepsilon^{\,}_{\bs{p},a}\,
\hat{c}^{\,}_{\bs{p},a}.
\end{equation}
\end{subequations}
The algebra~(\ref{eq: def fermion algebra lattice})
applies to the band operators labeled by the 
\textit{band index} $a$
if one multiplies the Kronecker symbol 
$\delta^{\,}_{\bs{p},\bs{p}'}$
in the Brillouin zone
by the Kronecker symbol
$\delta^{\,}_{a,a'}$ among the bands.
The algebra~(\ref{eq: def fermion algebra lattice}) thus
endows the orbital creation and annihilation operators
with the canonical fermion algebra. 

The interacting Hamiltonian is defined by
\begin{subequations}
\label{eq: def lattice interacting H 0 and U}
\begin{equation}
\begin{split}
\hat{H}^{\,}_{U}:=&\,
\sum_{\bs{r},\bs{r}'\in\Lambda}
\sum_{\alpha,\alpha'}
\hat{\rho}^{\psi}_{\bs{r} ,\alpha }\,
U^{\alpha,\alpha'}_{\bs{r}-\bs{r}'}\,
\hat{\rho}^{\psi}_{\bs{r}',\alpha'}
\\
=&\,
\sum_{\bs{q}\in\Omega^{\,}_{\mathrm{BZ}}}
\sum_{\alpha,\alpha'}
\hat{\rho}^{\psi}_{+\bs{q} ,\alpha }\,
\tilde{U}^{\alpha,\alpha'}_{\bs{q}}\,
\hat{\rho}^{\psi}_{-\bs{q},\alpha'},
\end{split}
\end{equation}
with
\begin{equation}
\hat{\rho}^{\psi}_{\bs{r},\alpha}:=
\hat\psi^{\dag}_{\bs{r},\alpha}\,
\hat\psi^{\;}_{\bs{r},\alpha}
\label{eq: density orbital basis}
\end{equation}
the local density at site $\bs{r}\in\Lambda$ 
and for the orbital $\alpha$.
The corresponding Fourier transforms are
\begin{equation}
\begin{split}
&
\hat{\rho}^{\psi}_{\bs{q},\alpha}:=
\sum_{\bs{p}\in\Omega^{\,}_{\mathrm{BZ}}}
\hat\psi^{\dag}_{[\bs{q}+\bs{p}]^{\,}_{\mathrm{BZ}},\alpha}\,
\hat\psi^{\;}_{\bs{p},\alpha},
\\
&
\tilde{U}^{\alpha,\alpha'}_{\bs{q}}:=
\frac{1}{N}
\sum_{\bs{r}\in\Lambda}
e^{-\mathrm{i}\,\bs{q}\cdot\bs{r}}\,
U^{\alpha,\alpha'}_{\bs{r}}.
\end{split}
\end{equation}
\end{subequations}
For simplicity, we shall focus on orbital-independent
(density-density) interactions, in which case
\begin{equation}
U^{\alpha,\alpha'}_{\bs{r}}=
U^{\,}_{\bs{r}},
\quad 
\forall \alpha,\alpha^\prime
\;.
\end{equation}

Normal ordering is the operation by which all
creation operators are to be moved to
the left of the annihilation operators.
In the orbital basis, normal ordering results in
\begin{subequations}
\begin{equation}
\hat{H}^{\,}_{U}=
\hat{H}^{\prime\psi}_{U}\,
+\,
\hat{H}^{\prime\prime\psi}_{U}\;.
\end{equation}
The one-body Hamiltonian $\hat{H}^{\prime\psi}_{U}$, 
a consequence of the fermion algebra,
is proportional to the conserved number operator,
\begin{equation}
\hat{H}^{\prime\psi}_{U}:=
\sum_{\bs{r}\in\Lambda}
\sum_{\alpha}
U^{\;}_{\bs{0}}\,
\hat{\psi}^{\dag}_{\bs{r},\alpha}\,
\hat{\psi}^{\,  }_{\bs{r},\alpha}\equiv
U\,\hat{N},
\end{equation}
where we defined $U\equiv U^{\,}_{\bs{r}=\bs{0}}$.
The normal-ordered interaction 
$\hat{H}^{\prime\prime\psi}_{U}$ is
\begin{equation}
\hat{H}^{\prime\prime\psi}_{U}\equiv
\sum_{\bs{r},\bs{r}'\in\Lambda}
\sum_{\alpha,\alpha'}
U^{\ }_{\bs{r}-\bs{r}'}\,
\hat{\psi}^{\dag}_{\bs{r} ,\alpha}\,
\hat{\psi}^{\dag}_{\bs{r}',\alpha'}\,
\hat{\psi}^{\,  }_{\bs{r}',\alpha'}\,
\hat{\psi}^{\,  }_{\bs{r} ,\alpha}.
\end{equation}
\end{subequations}

\begin{widetext}
The one-body term induced by normal ordering is,
in the band basis,
\begin{subequations} 
\begin{equation}
\hat{H}^{\prime c}_{U}:=
U
\sum_{\bs{p}\in\Omega^{\,}_{\mathrm{BZ}}}
\sum_{a}
\hat{c}^{\dag}_{\bs{p},a}\,
\hat{c}^{\,  }_{\bs{p},a}
\equiv
U\,
\hat{N}.
\end{equation}
The normal-ordered interaction is, in the band basis,
\begin{equation}
\begin{split}
&
\hat{H}^{\prime\prime c}_{U}:=
\sum_{\bs{q} \in\Omega^{\,}_{\mathrm{BZ}}}
\sum_{\bs{p} \in\Omega^{\,}_{\mathrm{BZ}}}
\sum_{\bs{p}'\in\Omega^{\,}_{\mathrm{BZ}}}
\sum_{a}
\sum_{b}
\sum_{a'}
\sum_{b'}
V^{a,b;a',b'}_{\bs{q},\bs{p},\bs{p}'}\,
\hat{c}^{\dag}_{[+\bs{q}+\bs{p} ]^{\,}_{\mathrm{BZ}},a }\,
\hat{c}^{\dag}_{[-\bs{q}+\bs{p}']^{\,}_{\mathrm{BZ}},a'}\,
\hat{c}^{\,  }_{\bs{p}',b'}\,
\hat{c}^{\,  }_{\bs{p} ,b },
\\
&
V^{a,b;a',b'}_{\bs{q},\bs{p},\bs{p}'}:=
\tilde{U}^{\,}_{\bs{q}}
\sum_{\alpha,\alpha'}
u^{\alpha\,*}_{[+\bs{q}+\bs{p} ]^{\,}_{\mathrm{BZ}},a }\,
u^{\alpha   }_{\bs{p} ,b }\,
u^{\alpha'\,*}_{[-\bs{q}+\bs{p}']^{\,}_{\mathrm{BZ}},a'}\,
u^{\alpha'   }_{\bs{p}',b'}.
\end{split}
\end{equation}
\end{subequations}
\end{widetext}

In any subspace of the Fock space with a fixed number of particles,
normal ordering thus produces a rigid shift of all single-particle
energy eigenvalues of $\hat{H}^{\,}_{0}$. For any band $a$,
the width of the single-particle dispersion
$\varepsilon^{\,}_{a}$
is not affected by the normal ordering,
i.e., by adding to or subtracting from $\hat{H}^{\,}_{0}$
the operator $U\,\hat{N}$.
We are going to show that this needs not be true any longer if
we \textit{first} project Hamiltonian~(\ref{eq: generic lattice H})
onto band $\bar{a}$ and \textit{then} express the resulting
projected Hamiltonian
in terms of the generators~(\ref{eq: def Phi G q p}).

The projection of Hamiltonian~(\ref{eq: generic lattice H})
onto band $\bar{a}$ is
\begin{subequations}
\label{eq: generic lattice H projected to bar a}
\begin{equation}
\hat{H}^{\bar{a}}=
\hat{H}^{\bar{a}}_{0}
+
\hat{H}^{\bar{a}}_{U},
\label{eq: generic lattice H projected to bar a A}
\end{equation}
where the projected kinetic Hamiltonian is
\begin{equation}
\hat{H}^{\bar{a}}_{0}=
\sum_{\bs{p}\in\Omega^{\,}_{\mathrm{BZ}}}
\hat{c}^{\dag}_{\bs{p},\bar{a}}\,
\left(
\varepsilon^{\,}_{\bs{p},\bar{a}}
+
U
\right)
\hat{c}^{\,}_{\bs{p},\bar{a}},
\label{eq: generic lattice H projected to bar a B}
\end{equation}
while the projected interacting Hamiltonian is
\begin{widetext}
\begin{equation}
\begin{split}
&
\hat{H}^{\bar{a}}_{U}=
\sum_{\bs{q} \in\Omega^{\,}_{\mathrm{BZ}}}
\sum_{\bs{p} \in\Omega^{\,}_{\mathrm{BZ}}}
\sum_{\bs{p}'\in\Omega^{\,}_{\mathrm{BZ}}}
V^{\bar{a}}_{\bs{q},\bs{p},\bs{p}'}\,
\hat{c}^{\dag}_{[+\bs{q}+\bs{p} ]^{\,}_{\mathrm{BZ}},\bar{a}}\,
\hat{c}^{\dag}_{[-\bs{q}+\bs{p}']^{\,}_{\mathrm{BZ}},\bar{a}}\,
\hat{c}^{\,  }_{\bs{p}',\bar{a}}\,
\hat{c}^{\,  }_{\bs{p} ,\bar{a}},
\\
&
V^{\bar{a}}_{\bs{q},\bs{p},\bs{p}'}=
\tilde{U}^{\,}_{\bs{q}}
\sum_{\alpha,\alpha'}
u^{\alpha\,*}_{[+\bs{q}+\bs{p} ]^{\,}_{\mathrm{BZ}},\bar{a}}\,
u^{\alpha   }_{\bs{p} ,\bar{a}}\,
u^{\alpha'\,*}_{[-\bs{q}+\bs{p}']^{\,}_{\mathrm{BZ}},\bar{a}}\,
u^{\alpha'   }_{\bs{p}',\bar{a}}.
\end{split}
\label{eq: generic lattice H projected to bar a C}
\end{equation}
\end{widetext}
\end{subequations}

For the purpose of representing
the projection of Hamiltonian~(\ref{eq: generic lattice H})
onto band $\bar{a}$ by the magnetic density operators%
~(\ref{eq: def Phi G q p}), it is necessary to undo the
normal ordering in Eq.%
~(\ref{eq: generic lattice H projected to bar a C}).
In doing so, a second one-body term is produced,
\begin{subequations}
\label{eq: generic lattice Hprime projected to bar a}
\begin{equation}
\hat{H}^{\bar{a}}=
\hat{H}^{\prime\bar{a}}_{0}
+
\hat{H}^{\prime\bar{a}}_{U},
\label{eq: generic lattice H projected to bar a AA}
\end{equation}
where the projected kinetic Hamiltonian is
\begin{widetext}
\begin{equation}
\begin{split}
\hat{H}^{\prime\bar{a}}_{0}=&\,
\sum_{\bs{p}\in\Omega^{\,}_{\mathrm{BZ}}}
\left(
\varepsilon^{\,}_{\bs{p},\bar{a}}
+
U
\right)\,
\hat{c}^{\dag}_{\bs{p},\bar{a}}\,
\hat{c}^{\,}_{\bs{p},\bar{a}}
-
\sum_{\bs{p}\in\Omega^{\,}_{\mathrm{BZ}}}
\left(\sum_{\bs{q}\in\Omega^{\,}_{\mathrm{BZ}}}
V^{\bar{a}}_{\bs{q},[-\bs{q}+\bs{p}]^{\,}_{\mathrm{BZ}},\bs{p}}
\right)\;
\hat{c}^{\dag}_{\bs{p},\bar{a}}\,
\hat{c}^{\,}_{\bs{p},\bar{a}},
\end{split}
\label{eq: generic lattice H projected to bar a BB}
\end{equation}
while the projected interacting Hamiltonian is
\begin{equation}
\begin{split}
&
\hat{H}^{\prime\bar{a}}_{U}=
\sum_{\bs{q} \in\Omega^{\,}_{\mathrm{BZ}}}
\sum_{\bs{p} \in\Omega^{\,}_{\mathrm{BZ}}}
\sum_{\bs{p}'\in\Omega^{\,}_{\mathrm{BZ}}}
V^{\bar{a}}_{\bs{q},\bs{p},\bs{p}'}\,
\hat{c}^{\dag}_{[+\bs{q}+\bs{p} ]^{\,}_{\mathrm{BZ}},\bar{a}}\,
\hat{c}^{\,  }_{\bs{p} ,\bar{a}}\,
\hat{c}^{\dag}_{[-\bs{q}+\bs{p}']^{\,}_{\mathrm{BZ}},\bar{a}}\,
\hat{c}^{\,  }_{\bs{p}',\bar{a}}.
\end{split}
\label{eq: generic lattice H projected to bar a CC}
\end{equation}
\end{widetext}
\end{subequations}
Observe that had we first represented 
Eq.~(\ref{eq: def lattice interacting H 0 and U})
in the band basis, followed by the projection 
consisting of restricting all
the band indices to $\bar{a}$
prior to normal-ordering, then we would have
obtained Eq.~(\ref{eq: generic lattice Hprime projected to bar a}) 
upon normal ordering without the
second term on the right-hand side of 
Eq.~(\ref{eq: generic lattice H projected to bar a BB}).
The correct implementation of projection is to normal order first
and then to project, leading to
Eq.~(\ref{eq: generic lattice H projected to bar a BB}). 
Indeed, the order by which normal ordering is followed 
by restricting all band indices
to the projected ones corresponds to sandwiching 
the Hamiltonian by the projection operator onto a subset of bands. 
The reverse order by which the density operators is projected
onto a subset of bands followed by normal ordering 
corresponds to sandwiching first all density operators
by the projection operator onto a subset of bands and then
assembling a Hamiltonian out of these projected density operators.
As the projection operators do not commute with the
density operators, the order in which the operations of
normal ordering and projection are performed matters.

\begin{widetext}
We can now express the Hamiltonian in terms of the magnetic density
operators (the details are provided in
Appendix~\ref{appendix: Equation MAIN RESULT})
\begin{subequations}
\label{eq: MAIN RESULT}
\begin{equation}
\begin{split}
&
\hat{H}^{\prime\bar{a}}_{0}=
\sum_{\bs{G}\,\in\Lambda^{\star}}
\left(
\tilde{\varepsilon}^{\,}_{\bs{G}}
+
U\,\delta^{\;}_{\bs{G},0}
-
\sum_{\bs{q}}\;\tilde{U}^{\,}_{\bs{q}}\,
\tilde{h}^{\bs{G}}_{-\bs{q}}
\right)
\hat{\varrho}^{\bs{G}}_{\bs{0}},
\\
&
\hat{H}^{\prime\bar{a}}_{U}=
\sum_{\bs{q}\in\Omega^{\,}_{\mathrm{BZ}}}
\tilde{U}^{\,}_{\bs{q}}\;
\sum_{\bs{G},\bs{G}'\,\in\Lambda^{\star}}
\tilde{f}^{\bs{G} }_{\bs{q}}\,
\tilde{f}^{-\bs{G}'}_{-\bs{q}}\,
e^{
-\mathrm{i}\,
(
\Theta^{\bs{G}}_{+\bs{q}}
+
\Theta^{-\bs{G}'}_{-\bs{q}}
)
  }
\;
\hat{\varrho}^{\bs{G} }_{\bs{q}}\,
\hat{\varrho}^{-\bs{G}'}_{-\bs{q}},
\end{split}
\end{equation}
where
\begin{eqnarray}
\tilde{\varepsilon}^{\,}_{\bs{G}}
&=&
\frac{1}{N}\;
\sum_{\bs{p}} 
\varepsilon^{\,}_{\bs{p}}\;
e^{
-\mathrm{i}\,
\left(
\bs{G}\,\bs{\ast}\,\bs{p}
-
\bs{p}\,\bs{\ast}\,\bs{G}
\right)
/(2\pi)
  },
\\
\tilde{h}^{\bs{G}}_{\bs{q}}
&=&
\frac{1}{N}\;
\sum_{\bs{p}} 
\left|
\sum_\alpha u^{\alpha\,*}_{[\bs{q}+\bs{p}]^{\,}_{\mathrm{BZ}},\bar a}\,
u^{\alpha   }_{\bs{p},\bar a}
\right|^2
\;
e^{
-\mathrm{i}\,
\left(
\bs{G}\,\bs{\ast}\,\bs{p}
-
\bs{p}\,\bs{\ast}\,\bs{G}
\right)
/(2\pi)
  },
\\
\tilde{f}^{\bs{G}}_{\bs{q}}
&=&
\frac{1}{N}\;
\sum_{\bs{p}} 
\left(
\sum_\alpha u^{\alpha\,*}_{[\bs{q}+\bs{p}]^{\,}_{\mathrm{BZ}},\bar a}\,
u^{\alpha   }_{\bs{p},\bar a}
\right)
\;
e^{-\mathrm{i}\,\Theta^{\,}_{\bs{q},\bs{p}}}\,
e^{
-\mathrm{i}\,
\left(
\bs{G}\,\bs{\ast}\,\bs{p}
-
\bs{p}\,\bs{\ast}\,\bs{G}
\right)
/(2\pi)
  },
\end{eqnarray}
\end{subequations}
with $\Theta^{\,}_{\bs{q},\bs{p}}$ and $\Theta^{\bs{G}}_{\bs{q}}$
defined in Eqs.~(\ref{eq: evaluation Phi q and p lattice a})
and~(\ref{eq: evaluation Phi q and p lattice b}), respectively.
\end{widetext}

Equation~(\ref{eq: MAIN RESULT}) is the main result
of Sec.~\ref{subsec: Normal ordering and the bare band width}.
Applied to a Chern insulator to which density-density
interactions have been added, 
Eq.~(\ref{eq: MAIN RESULT})
suggests that there will always be linear in 
$\hat{\varrho}^{\bs{G}}_{\bs{q=0}}$ 
contributions to the Hamiltonian even if the bare band is flat
to begin with, i.e., even if $\varepsilon^{\,}_{\bs{p},\bar{a}}=0$. 
Because of the topological attributes of the Bloch spinors 
as they wrap around the Brillouin zone, we expect
a nonvanishing $\tilde{h}^{\bs{G}}_{\bs{q}}$.
(An extreme case of a topologically trivial band insulator
has Bloch spinors that are constant across the Brillouin zone,
in which case only $\tilde{h}^{\bs{G=0}}_{\bs{q}}\ne 0$ and 
the additional one-body contribution
is just proportional to the total particle number.
This would also be the case in the context of the quantum Hall effect.) 
This effect on the bare dispersion is controlled by the
bare interaction $U^{\,}_{\bs{q}}$.
Hence, it could be as large as the effects of the
density-density interaction.

It is far from evident that
a FCI is selected by the competition between the
one-body and two-body terms in
Eq.~(\ref{eq: MAIN RESULT})
since they are both controlled by one characteristic energy
scale in the limit of a flat bare bandwidth.
On the other hand, if a ground state supporting a FCI
is selected for some range of parameters, then the
effective quantum field theory
describing the low-energy and long-distance properties of this
phase should belong to one of the
universality class associated with the FQHE. 

\section*{Acknowledgments}

This work was supported in part by DOE Grant No.\ DEFG02-06ER46316.

\appendix

\section{
Proof of Eq.~(\ref{eq: closed GMP algebra on the lattice})
        }
\label{appsec: Proof closure algebra}

For any $\bs{q}$ and $\bs{q}'$ 
from the first Brillouin zone $\Omega^{\,}_{\mathrm{BZ}}$
and for any $\bs{G}$ and $\bs{G}'$ 
from the dual lattice $\Lambda^{\star}$,
Eq.~(\ref{eq: commutator two rhos}) 
dictates that
\begin{subequations}
\begin{widetext}
\begin{equation}
\begin{split}
\left[
\hat{\varrho}^{\bs{G} }_{\bs{q} },
\hat{\varrho}^{\bs{G}'}_{\bs{q}'}
\right]=&\,
\sum_{\bs{p}\,\in\Omega^{\,}_{\mathrm{BZ}}}
\left[
e^{
+\mathrm{i}\,\Phi^{\bs{G}}_{\bs{q},[\bs{q}'+\bs{p}]^{\,}_{\mathrm{BZ}}}
+\mathrm{i}\,\Phi^{\bs{G}'}_{\bs{q}',\bs{p}}
  }
-
(
\bs{q}\leftrightarrow\bs{q}'
\hbox{ and }
\bs{G}\leftrightarrow\bs{G}'
)
\right]\,
\hat{c}^{\dag}_{
[\bs{q}+\bs{q}'+\bs{p}]^{\,}_{\mathrm{BZ}}
               }\,
\hat{c}^{\,  }_{\bs{p} }
\\
=&\,
\sum\limits_{\bs{p}\,\in\Omega^{\,}_{\mathrm{BZ}}}
e^{
+\mathrm{i}\,
\Phi^{\bs{G}''}_{[\bs{q}+\bs{q}']^{\,}_{\mathrm{BZ}},\bs{p}}
  }
\left[
e^{
+\mathrm{i}\,
\Phi^{\bs{G}}_{\bs{q},[\bs{q}'+\bs{p}]^{\,}_{\mathrm{BZ}}}
+\mathrm{i}\,
\Phi^{\bs{G}'}_{\bs{q}',\bs{p}}
-\mathrm{i}\,
\Phi^{\bs{G}''}_{[\bs{q}+\bs{q}']^{\,}_{\mathrm{BZ}},\bs{p}}
  }
-
(
\bs{q}\leftrightarrow\bs{q}'
\hbox{ and }
\bs{G}\leftrightarrow\bs{G}'
)
\right]\,
\hat{c}^{\dag}_{
[\bs{q}+\bs{q}'+\bs{p}]^{\,}_{\mathrm{BZ}}
              }\,
\hat{c}^{\,  }_{\bs{p} }
\\
\equiv&\,
\sum\limits_{\bs{p}\,\in\Omega^{\,}_{\mathrm{BZ}}}
\mathcal{F}^{\bs{G} ,\bs{G}',\bs{G}''}_{\bs{q} ,\bs{q}',\bs{p}}\
e^{
+\mathrm{i}\,
\Phi^{\bs{G}''}_{[\bs{q}+\bs{q}',\bs{p}]^{\,}_{\mathrm{BZ}}}
  }\
\hat{c}^{\dag}_{
[\bs{q}+\bs{q}'+\bs{p}]^{\,}_{\mathrm{BZ}}
              }\,
\hat{c}^{\,  }_{\bs{p} }.
\end{split}
\label{appeq: commutators 2 rhoG's}
\end{equation}
\end{widetext}
Here, we have introduced the short-hand notations
\begin{equation}
[\bs{q}+\bs{p}]^{\,}_{\mathrm{BZ}}\equiv
\bs{q}+\bs{p}-\bs{G}^{\,}_{\bs{q}+\bs{p}}
\,\in\Omega^{\,}_{\mathrm{BZ}},
\label{appeq: def notation [...]BZ}
\end{equation}
\begin{equation}
\mathcal{F}^{\bs{G} ,\bs{G}',\bs{G}''}_{\bs{q} ,\bs{q}',\bs{p}}:=
e^{
+\mathrm{i}\,
\Gamma^{\bs{G},\bs{G}',\bs{G}''}_{\bs{q},\bs{q}',\bs{p}}
  }
-
(
\bs{q}\leftrightarrow\bs{q}'
\hbox{ and }
\bs{G}\leftrightarrow\bs{G}'
),
\end{equation}
and
\begin{equation}
\Gamma^{\bs{G},\bs{G}',\bs{G}''}_{\bs{q},\bs{q}',\bs{p}}:=
\Phi^{\bs{G}}_{\bs{q},[\bs{q}'+\bs{p}]^{\,}_{\mathrm{BZ}}}
+
\Phi^{\bs{G}'}_{\bs{q}',\bs{p}}
-
\Phi^{\bs{G}''}_{[\bs{q}+\bs{q}']^{\,}_{\mathrm{BZ}},\bs{p}}
\label{appeq: def Upsilon G G' G'' q q' p}
\end{equation}
\end{subequations}
for any triplet $\bs{q}$, $\bs{q}'$, $\bs{p}$  
from the first Brillouin zone and
for any triplet $\bs{G}$, $\bs{G}'$, $\bs{G}''$  
from the dual lattice.

First, we are going to show that
the algebra~(\ref{appeq: commutators 2 rhoG's})
closes to
\begin{subequations}
\begin{equation}
\left[
\hat{\varrho}^{\bs{G} }_{\bs{q} },
\hat{\varrho}^{\bs{G}'}_{\bs{q}'}
\right]=
F^{\bs{G} ,\bs{G}'}_{\bs{q} ,\bs{q}'}\,
\hat{\varrho}^{\bs{G} +\bs{G}'}_{\bs{q}+\bs{q}'}
\label{appeq: closed algebra due to no p dependence}
\end{equation}
as a consequence of the fact that the kernel
$\mathcal{F}^{\bs{G} ,\bs{G}',\bs{G}''}_{\bs{q} ,\bs{q}',\bs{p}}$
is independent of $\bs{p}$ and $\bs{G}''$
on the right-hand side of Eq.~(\ref{appeq: commutators 2 rhoG's})
(as we will show shortly)
and for which reason we have introduced the notations
\begin{equation}
\Gamma^{\bs{G},\bs{G}',\bs{G}''}_{\bs{q},\bs{q}',\bs{p}}\equiv
\Upsilon^{\bs{G},\bs{G}'}_{\bs{q},\bs{q}'}
\qquad
\mathcal{F}^{\bs{G} ,\bs{G}',\bs{G}''}_{\bs{q} ,\bs{q}',\bs{p}}\equiv
F^{\bs{G} ,\bs{G}'}_{\bs{q} ,\bs{q}'}.
\label{appeq: p falls out Upsilon}
\end{equation}
\end{subequations}
Second, we are going to show that
the algebra~(\ref{appeq: closed algebra due to no p dependence})
simplifies to the algebra
\begin{subequations}
\begin{equation}
\left[
\hat{\varrho}^{\bs{G} }_{\bs{q} },
\hat{\varrho}^{\bs{G}'}_{\bs{q}'}
\right]=
2\mathrm{i}\,
\sin\,
\left(
\frac{
(\bs{q}+\bs{G})\,\bs{\ast}\,(\bs{q}'+\bs{G}')
     }
     {
2\pi
     }
\right)\,
\hat{\varrho}^{\bs{G} +\bs{G}'}_{\bs{q}+\bs{q}'}
\label{appeq: closed algebra due to no p dependence and as}
\end{equation}
as a consequence of the fact that (as we will show shortly)
\begin{equation}
\Upsilon^{\bs{G},\bs{G}'}_{\bs{q},\bs{q}'}=
-
\Upsilon^{\bs{G}',\bs{G}}_{\bs{q}',\bs{q}}.
\label{appeq: Upsilon is as}
\end{equation}
\end{subequations}

\begin{widetext}
\begin{proof}

Equations~(\ref{appeq: p falls out Upsilon})
and (\ref{appeq: Upsilon is as})
follow at once from choosing
\begin{subequations}
\begin{equation}
\Phi^{\bs{G}}_{\bs{q},\bs{p}}:=
\frac{1}{2\pi}
\left[
(\bs{q}+\bs{G})\bs{\ast}\bs{p}
+
\varphi^{\bs{G}}_{\bs{q},\bs{p}}
\right],
\label{appeq: def Phi G q p}
\end{equation}
where
\begin{equation}
\varphi^{\bs{G}}_{\bs{q},\bs{p}}:=
-
(\bs{q}+\bs{p}+\bs{G})
\bs{\ast}
\bs{G}^{\,}_{\bs{q}+\bs{p}+\bs{G}}.
\label{appeq: def varphi G q p}
\end{equation}
\end{subequations}

To verify this claim, we start from
\begin{subequations}
\begin{equation}
\begin{split}
2\pi\,
\Gamma^{\bs{G},\bs{G}',\bs{G}''}_{\bs{q},\bs{q}',\bs{p}}=&\,
2\pi
\left(
\Phi^{\bs{G}}_{\bs{q},[\bs{q}'+\bs{p}]^{\,}_{\mathrm{BZ}}}
+
\Phi^{\bs{G}'}_{\bs{q}',\bs{p}}
-
\Phi^{\bs{G}''}_{[\bs{q}+\bs{q}']^{\,}_{\mathrm{BZ}},\bs{p}}
\right)
\\
=&\,
(\bs{q}+\bs{G})\bs{\ast}[\bs{q}'+\bs{p}]^{\,}_{\mathrm{BZ}}
+
(\bs{q}'+\bs{G}')\bs{\ast}\bs{p}
-
([\bs{q}+\bs{q}']^{\,}_{\mathrm{BZ}}+\bs{G}'')\bs{\ast}\bs{p}
+
2\pi\,\Xi^{\bs{G},\bs{G}',\bs{G}''}
_{\bs{q},\bs{q}',\bs{p}},
\end{split}
\end{equation}
where
\begin{equation}
\begin{split}
2\pi\,\Xi^{\bs{G},\bs{G}',\bs{G}''}
_{\bs{q},\bs{q}',\bs{p}}:=&\,
\varphi^{\bs{G}}_{\bs{q},[\bs{q}'+\bs{p}]^{\,}_{\mathrm{BZ}}}
+
\varphi^{\bs{G}'}_{\bs{q}',\bs{p}}
-
\varphi^{\bs{G}''}_{[\bs{q}+\bs{q}']^{\,}_{\mathrm{BZ}},\bs{p}}.
\end{split}
\label{eq: def Xi  GG'G''qq'p a}
\end{equation}
\end{subequations}
If it were not for the symbol $[\cdots]^{\,}_{\mathrm{BZ}}$
and the dependence on $\bs{G}$, $\bs{G}'$, and $\bs{G}''$,
all the \textit{explicit} dependence on $\bs{p}$ 
would drop by linearity on the right-hand side,
very much as was the case in the continuum for Eq.%
~(\ref{eq: evaluation Upsilon continuum}).
The condition for
periodicity prevents this cancellation, however. Instead,
according to Eq.~(\ref{eq: def noation [...]BZ}),
\begin{equation}
\begin{split}
2\pi\,
\Gamma^{\bs{G},\bs{G}',\bs{G}''}_{\bs{q},\bs{q}',\bs{p}}=&\,
(
\bs{q}
+
\bs{G}
)
\bs{\ast}
(
\underline{
\bs{q}'
          }
+
\bs{p}
-
\underline{\underline{
\bs{G}^{\,}_{\bs{q}'+\bs{p}}
          }          }
)
+
(\bs{q}'+\bs{G}')\bs{\ast}\bs{p}
-
(\bs{q}+\bs{q}'-\bs{G}^{\,}_{\bs{q}+\bs{q}'}+\bs{G}'')\bs{\ast}\bs{p}
+
2\pi\,\Xi^{\bs{G},\bs{G}',\bs{G}''}
_{\bs{q},\bs{q}',\bs{p}}
\end{split}
\end{equation}
simplifies to, if we collect all terms explicitly linear in $\bs{p}$,
\begin{equation}
\begin{split}
2\pi\,
\Gamma^{\bs{G},\bs{G}',\bs{G}''}_{\bs{q},\bs{q}',\bs{p}}=&\,
(\bs{q}+\bs{G})
\bs{\ast}
\underline{\bs{q}'}
-
(\bs{q}+\bs{G})
\bs{\ast}
\underline{\underline{
\bs{G}^{\,}_{\bs{q}'+\bs{p}}
          }          }
+
(\bs{G}+\bs{G}'+\bs{G}^{\,}_{\bs{q}+\bs{q}'}-\bs{G}'')\bs{\ast}\bs{p}
+
2\pi\,\Xi^{\bs{G},\bs{G}',\bs{G}''}
_{\bs{q},\bs{q}',\bs{p}}.
\end{split}
\end{equation}
We choose 
\begin{equation}
\bs{G}''=
\bs{G}+\bs{G}'+\bs{G}^{\,}_{\bs{q}+\bs{q}'}.
\label{eq: clever choice of G''}
\end{equation} 
Then, all terms explicitly linear in $\bs{p}$ drop out 
and we are left with
\begin{equation}
\begin{split}
2\pi\,
\Gamma^{\bs{G},\bs{G}',\bs{G}+\bs{G}'+\bs{G}^{\,}_{\bs{q}+\bs{q}'}}
_{\bs{q},\bs{q}',\bs{p}}=&\,
+
(\bs{q}+\bs{G})\bs{\ast}\bs{q}'
-
\underline{\underline{
(\bs{q}+\bs{G})\bs{\ast}\bs{G}^{\,}_{\bs{q}'+\bs{p}}
          }          }
+
2\pi\,\Xi^{\bs{G},\bs{G}',\bs{G}+\bs{G}'+\bs{G}^{\,}_{\bs{q}+\bs{q}'}}
_{\bs{q},\bs{q}',\bs{p}}.
\end{split}
\label{eq: intermediary Upsilon G G' G'' q q' p}
\end{equation}
An implicit dependence on $\bs{p}$ remains through the underlined term,
on the one hand, and the functions of the form
$\varphi^{\bs{G}}_{\bs{q},\bs{p}}$,
on the other hand.

It is time to evaluate the contribution
$
2\pi\,\Xi^{\bs{G},\bs{G}',\bs{G}''}
_{\bs{q},\bs{q}',\bs{p}}
$.
Observe that
\begin{equation}
\begin{split}
\bs{G}^{\,}_{\bs{q}+\bs{p}+\bs{G}}=&\,
\bs{q}+\bs{p}+\bs{G}
-
[\bs{q}+\bs{p}+\bs{G}]^{\,}_{\mathrm{BZ}}
\\
=&\,
\bs{q}+\bs{p}+\bs{G}
-
(\bs{q}+\bs{p}+\bs{G}-\bs{G}^{\,}_{\bs{q}+\bs{p}}-\bs{G})
\\
=&\,
\bs{G}^{\,}_{\bs{q}+\bs{p}}
+
\bs{G}
\end{split}
\end{equation}
for any pair $\bs{q}$ and $\bs{p}$ from the Brillouin zone 
$\Omega^{\,}_{\mathrm{BZ}}$ and for any $\bs{G}$ from the dual lattice
$\Lambda^{\star}$.
Hence, we can rewrite
\begin{equation}
\begin{split}
2\pi\,
\Xi^{\bs{G},\bs{G}',\bs{G}''}
_{\bs{q},\bs{q}',\bs{p}}=&\,
-
(\bs{q}+\bs{q}'+\bs{p}-\bs{G}^{\,}_{\bs{q}'+\bs{p}}+\bs{G})
\bs{\ast}
\bs{G}^{\,}_{\bs{q}+\bs{q}'+\bs{p}-\bs{G}^{\,}_{\bs{q}'+\bs{p}}+\bs{G}}
\\
&\,
-
(\bs{q}'+\bs{p}+\bs{G}')
\bs{\ast}
\bs{G}^{\,}_{\bs{q}'+\bs{p}+\bs{G}'}
\\
&\,
+
(\bs{q}+\bs{q}'+\bs{p}-\bs{G}^{\,}_{\bs{q}+\bs{q}'}+\bs{G}'')
\bs{\ast}
\bs{G}^{\,}_{\bs{q}+\bs{q}'+\bs{p}-\bs{G}^{\,}_{\bs{q}+\bs{q}'}+\bs{G}''}
\end{split}
\label{eq: intermediary step for XI G G' G'' q q' p} 
\end{equation}
as
\begin{equation}
\begin{split}
2\pi\,\Xi^{\bs{G},\bs{G}',\bs{G}''}
_{\bs{q},\bs{q}',\bs{p}}=&\,
-
(\bs{q}+\bs{q}'+\bs{p}-\bs{G}^{\,}_{\bs{q}'+\bs{p}}+\bs{G})
\bs{\ast}
(\bs{G}^{\,}_{\bs{q}+\bs{q}'+\bs{p}}-\bs{G}^{\,}_{\bs{q}'+\bs{p}}+\bs{G})
\\
&\,
-
(\bs{q}'+\bs{p}+\bs{G}')
\bs{\ast}
(\bs{G}^{\,}_{\bs{q}'+\bs{p}}+\bs{G}')
\\
&\,
+
(\bs{q}+\bs{q}'+\bs{p}-\bs{G}^{\,}_{\bs{q}+\bs{q}'}+\bs{G}'')
\bs{\ast}
(\bs{G}^{\,}_{\bs{q}+\bs{q}'+\bs{p}}-\bs{G}^{\,}_{\bs{q}+\bs{q}'}+\bs{G}'').
\end{split}
\label{eq: intermediary step for XI G G' G'' q q' p bis} 
\end{equation}
If we use Eq.%
~(\ref{eq: clever choice of G''})
on the last line of Eq.%
~(\ref{eq: intermediary step for XI G G' G'' q q' p bis}), 
we get
\begin{equation}
\begin{split}
2\pi\,\Xi^{\bs{G},\bs{G}',\bs{G}+\bs{G}'+\bs{G}^{\,}_{\bs{q}+\bs{q}'}}
_{\bs{q},\bs{q}',\bs{p}}=&\,
-
\left[
(
\bs{q}
+
\bs{q}'
+
\bs{p}
)
-
(
\bs{G}^{\,}_{\bs{q}'+\bs{p}}
-
\bs{G}
\hphantom{
+
\bs{G}'
         }\ \,
)
\right]
\bs{\ast}
(
\underline{
\bs{G}^{\,}_{\bs{q}+\bs{q}'+\bs{p}}
          }_{\#1}
-
\underline{
\bs{G}^{\,}_{\bs{q}'+\bs{p}}
          }_{\#2}
+
\underline{
\bs{G}
          }_{\#3}
)
\\
&\,
-
\left[
(
\hphantom{
\bs{q}+\ \,
         }
\bs{q}'
+
\bs{p}
)
-
(
\hphantom{
\bs{G}^{\,}_{\bs{q}'+\bs{p}}
+
\bs{G}
         }
-
\bs{G}'
)
\right]
\bs{\ast}
(
\underline{
\bs{G}^{\,}_{\bs{q}'+\bs{p}}
          }_{\#2}
+
\underline{
\bs{G}'
          }_{\#4}
)
\\
&\,
+
\left[
(
\bs{q}
+
\bs{q}'
+
\bs{p}
)
-
(
\hphantom{
\bs{G}^{\,}_{\bs{q}'+\bs{p}}
         }
-
\bs{G}
-
\bs{G}'
)
\right]
\bs{\ast}
(
\underline{
\bs{G}^{\,}_{\bs{q}+\bs{q}'+\bs{p}}
          }_{\#1}
+
\underline{
\bs{G}
          }_{\#3}
+
\underline{
\bs{G}'   
          }_{\#4}
).
\end{split}
\label{eq: intermediary step for XI G G' G'' q q' p bis bis} 
\end{equation} 
The following terms cancel pairwise.
\begin{enumerate}
\item From lines 1 and 3 
on the right-hand side of 
Eq.~(\ref{eq: intermediary step for XI G G' G'' q q' p bis bis}),
$
(\bs{q}+\bs{q}'+\bs{p}+\bs{G})\bs{\ast}
\underline{
\bs{G}^{\,}_{\bs{q}+\bs{q}'+\bs{p}}
          }_{\#1}
$
cancel;
$
(\bs{G}^{\,}_{\bs{q}'+\bs{p}}+\bs{G}')\bs{\ast}
\bs{G}^{\,}_{\bs{q}+\bs{q}'+\bs{p}}
$
is left over.
\item From lines 1 and 2 on the right-hand side of 
Eq.~(\ref{eq: intermediary step for XI G G' G'' q q' p bis bis}),
$
(\hphantom{\bs{q}+\ }\bs{q}'+\bs{p})\bs{\ast}
\underline{
\bs{G}^{\,}_{\bs{q}'+\bs{p}}
          }_{\#2}
$
cancel;
$
(
\underline{\underline{
\bs{q}
          }          }
-
\bs{G}^{\,}_{\bs{q}'+\bs{p}}
+
\underline{\underline{
\bs{G}
          }          }
+
\bs{G}'
)
\bs{\ast}
\bs{G}^{\,}_{\bs{q}'+\bs{p}}
$
is left over.
\item From lines 1 and 3 on the right-hand side of 
Eq.~(\ref{eq: intermediary step for XI G G' G'' q q' p bis bis}),
$
(\bs{q}+\bs{q}'+\bs{p}+\bs{G})\bs{\ast}
\underline{
\bs{G}
          }_{\#3}
$
cancel;
$
(
\bs{G}^{\,}_{\bs{q}'+\bs{p}}
+
\bs{G}'
)
\bs{\ast}\bs{G}
$
is left over.
\item From lines 2 and 3 on the right-hand side of 
Eq.~(\ref{eq: intermediary step for XI G G' G'' q q' p bis bis}),
$
(\bs{q}'+\bs{p}+\bs{G}')\bs{\ast}
\underline{
\bs{G}'
          }_{\#4}
$
cancel;
$
(\bs{q}+\bs{G})\bs{\ast}\bs{G}'
$
is left over.
\end{enumerate}
Collecting all nonvanishing contributions
on the right-hand side of
Eq.%
~(\ref{eq: intermediary step for XI G G' G'' q q' p bis bis})
yields
\begin{equation}
\begin{split}
2\pi\,
\Xi^{\bs{G},\bs{G}',\bs{G}+\bs{G}'+\bs{G}^{\,}_{\bs{q}+\bs{q}'}}
_{\bs{q},\bs{q}',\bs{p}}=&\,
(\bs{G}^{\,}_{\bs{q}'+\bs{p}}+\bs{G}')\bs{\ast}\bs{G}^{\,}_{\bs{q}+\bs{q}'+\bs{p}}
-
(
\bs{G}^{\,}_{\bs{q}'+\bs{p}}
-
\bs{G}'
)
\bs{\ast}
\bs{G}^{\,}_{\bs{q}'+\bs{p}}
+
(
\bs{G}^{\,}_{\bs{q}'+\bs{p}}
+
\bs{G}'
)
\bs{\ast}\bs{G}
\\
&\,
+
\underline{\underline{
(\bs{q}+\bs{G})
          }          }
\bs{\ast}
\bs{G}^{\,}_{\bs{q}'+\bs{p}}
+
(\bs{q}+\bs{G})\bs{\ast}\bs{G}'.
\end{split}
\label{eq: final Xi G G' G''}
\end{equation}

By assumption~(\ref{eq: * product set by Bravais lattice})
the $\bs{\ast}$ product between any pair from the dual lattice
$\Lambda^{\star}$ is a multiple of $(2\pi)^{2}$.
Hence, combining Eq.~(\ref{eq: final Xi G G' G''})
with Eq.~(\ref{eq: intermediary Upsilon G G' G'' q q' p})
delivers the desired expression
\begin{equation}
\begin{split}
\Gamma^{\bs{G},\bs{G}',\bs{G}+\bs{G}'+\bs{G}^{\,}_{\bs{q}+\bs{q}'}}
_{\bs{q},\bs{q}',\bs{p}}=&\,
\frac{1}{2\pi}
(\bs{q}+\bs{G})\bs{\ast}(\bs{q}'+\bs{G}')
+
\mathrm{mod}\,2\pi
\equiv
\Upsilon^{\bs{G},\bs{G}'}_{\bs{q},\bs{q}'}=
-
\Upsilon^{\bs{G}'\bs{G} }_{\bs{q}',\bs{q} }.
\end{split}
\end{equation}

\end{proof}
\end{widetext}

\section{
Details for reaching Eq.~(\ref{eq: MAIN RESULT})
        }
\label{appendix: Equation MAIN RESULT}

Equipped with 
Eq.~(\ref{eq: generic lattice Hprime projected to bar a}),
we are in position to take advantage of the fact that,
for any even dimension $d$, the magnetic density operators%
~(\ref{eq: def Phi G q p}) form a complete basis for the
charge-neutral fermion bilinears made out of the band
creation and annihilation operators.
Define the functions
$
f^{i}:
\Omega^{\,}_{\mathrm{BZ}}\times\Omega^{\,}_{\mathrm{BZ}}
\to
\mathbb{C},
$
with $i=1,2,3,4$ by
\begin{subequations}
\begin{equation}
\begin{split}
&
f^{1}_{\bs{q},\bs{p}}:=
\delta^{\,}_{\bs{q},\bs{0}}\,
\varepsilon^{\,}_{\bs{p},\bar{a}},
\\
&
f^{2}_{\bs{q},\bs{p}}:=
\delta^{\,}_{\bs{q},\bs{0}}\, 
U,
\\
&
f^{3}_{\bs{q}',\bs{p}'}:=
\delta^{\,}_{\bs{q}',\bs{0}}\, 
\delta^{\,}_{\bs{p}',\bs{p}},
\\
&
f^{4}_{\bs{q},\bs{p}}:=
u^{\alpha\,*}_{[\bs{q}+\bs{p} ]^{\,}_{\mathrm{BZ}},\bar{a}}\,
u^{\alpha   }_{\bs{p} ,\bar{a}}.
\end{split}
\end{equation}
(Notice that $\bs{p}$ is a parameter in the definition of $f^{3}$.)
In terms of the operators defined in
Eq.~(\ref{eq: def varrho lattice}),
the projected kinetic Hamiltonian is
\begin{equation}
\hat{H}^{\prime\bar{a}}_{0}=
\sum_{\bs{q}\in\Omega^{\,}_{\mathrm{BZ}}}
\left(
\hat{\varrho}^{f^{1}}_{\bs{q}}
+
\hat{\varrho}^{f^{2}}_{\bs{q}}
-
\sum_{\bs{p},\bs{q}'\in\Omega^{\,}_{\mathrm{BZ}}}
V^{\bar{a}}_{\bs{q},[-\bs{q}+\bs{p}]^{\,}_{\mathrm{BZ}},\bs{p}}\;
\hat{\varrho}^{f^{3}}_{\bs{q}'}
\right),
\label{eq: projected kin varrho}
\end{equation}
while the projected interacting Hamiltonian is
\begin{equation}
\hat{H}^{\prime\bar{a}}_{U}=
\sum_{\bs{q}\in\Omega^{\,}_{\mathrm{BZ}}}
\tilde U^{\,}_{\bs{q}}\,
\hat{\varrho}^{f^{4}}_{\bs{q}}\,
\hat{\varrho}^{f^{4}}_{-\bs{q}}.
\label{eq: projected int varrho}
\end{equation}
\end{subequations}
The bare kinetic energy
is the first term on the right-hand side of
Eq.~(\ref{eq: projected kin varrho}).
The correction to the bare kinetic energy
from standard normal ordering
is the second term on the right-hand side of
Eq.~(\ref{eq: projected kin varrho}).
The function $f^{2}$ is a $\delta$ function
with respect to its first argument and constant with
respect to its second argument. This correction
does not change the bandwidth of the bare dispersion.
The last correction to the bare kinetic energy
is the third term on the right-hand side of
Eq.~(\ref{eq: projected kin varrho}).
It is controlled by the interaction $\tilde U^{\,}_{\bs{q}}$
dressed by the Bloch functions
that diagonalize the bare kinetic energy,
\begin{equation}
V^{\bar{a}}_{\bs{q},[-\bs{q}+\bs{p}]^{\,}_{\mathrm{BZ}},\bs{p}}=
\tilde U^{\,}_{\bs{q}}
\left|
\sum_{\alpha}
u^{\alpha\,*}_{[-\bs{q}+\bs{p}]^{\,}_{\mathrm{BZ}},\bar{a}}\,
u^{\alpha   }_{\bs{p},\bar{a}}
\right|^{2}.
\end{equation}
It can change the bare bandwidth by 
an amount of the order of the characteristic
energy of the interaction.

Finally, Eq.~(\ref{eq: from f to tilde f lattice}) 
dictates that we define the functions
$
\bar{f}^{i}:
\Omega^{\,}_{\mathrm{BZ}}\times\Omega^{\,}_{\mathrm{BZ}}
\to
\mathbb{C},
$
with $i=1,2,3,4$ by
\begin{equation}
\bar{f}^{i}_{\bs{q},\bs{p}}:=
e^{-\mathrm{i}\,\Theta^{\,}_{\bs{q},\bs{p}}}\,
f^{i}_{\bs{q},\bs{p}},
\end{equation}
where the argument of the multiplicative 
exponential factor on the right-hand side
was defined in
Eq.~(\ref{eq: evaluation Phi q and p lattice b}).
We observe that any one of the three functions 
$f^{1}$,
$f^{2}$,
and
$f^{3}$
is proportional to $\delta^{\,}_{\bs{q},\bs{0}}$
so that Eq.%
~(\ref{eq: Theta-q-p and Theta-q-G can vanish if q=0})
implies that
\begin{equation}
\bar{f}^{i}_{\bs{q},\bs{p}}:=
\delta^{\,}_{\bs{q},\bs{0}}\,
f^{i}_{\bs{0},\bs{p}},
\qquad
i=1,2,3.
\end{equation}
In even dimensional space, we can safely use the
$\bs{\ast}$-Fourier expansion 
[see Eq.~(\ref{eq: Fourier trsf lattice})]
\begin{equation}
\begin{split}
&
\bar{f}^{i}_{\bs{q},\bs{p}}=
\delta^{\,}_{\bs{q},\bs{0}}\,
\sum_{\bs{G}\,\in\Lambda^{\star}}
\tilde{f}^{i;\bs{G}}_{\bs{0}}\,
e^{
+\mathrm{i}\,
\left(
\bs{G}\,\bs{\ast}\,\bs{p}
-
\bs{p}\,\bs{\ast}\,\bs{G}
\right)
/(2\pi)
  },
\quad
i=1,2,3,
\\
&
\bar{f}^{4}_{\bs{q},\bs{p}}=
\sum_{\bs{G}\,\in\Lambda^{\star}}
\tilde{f}^{4;\bs{G}}_{\bs{q}}\,
e^{
+\mathrm{i}\,
\left(
\bs{G}\,\bs{\ast}\,\bs{p}
-
\bs{p}\,\bs{\ast}\,\bs{G}
\right)
/(2\pi)
  },
\\
&
\sum_{\alpha}
\left|
u^{\alpha\,*}_{[-\bs{q}+\bs{p}]^{\,}_{\mathrm{BZ}},\bar{a}}\,
u^{\alpha   }_{\bs{p},\bar{a}}
\right|^{2}=
\sum_{\bs{G}\,\in\Lambda^{\star}}
\tilde{h}^{-\bs{G}}_{-\bs{q}}\,
e^{
+\mathrm{i}\,
\left(
\bs{G}\,\bs{\ast}\,\bs{p}
-
\bs{p}\,\bs{\ast}\,\bs{G}
\right)
/(2\pi)
  },
\end{split}
\end{equation}
to compute the Fourier coefficients
$\tilde{f}^{i;\bs{G}}_{\bs{q}}$ with $i=1,2,3$,
$\tilde{f}^{4;\bs{G}}_{\bs{q}}$,
and $\tilde{h}^{\bs{G}}_{\bs{q}}$.
Application of
Eq.~(\ref{eq: eq stating completeness})
then delivers the desired representation
of the projection of Hamiltonian~(\ref{eq: generic lattice H})
onto the band $\bar{a}$ by the magnetic density operators%
~(\ref{eq: def Phi G q p}),
\begin{equation}
\begin{split}
&
\hat{H}^{\bar{a}}=
\hat{H}^{\prime\bar{a}}_{0}
+
\hat{H}^{\prime\bar{a}}_{U},
\\
&
\hat{H}^{\prime\bar{a}}_{0}=
\sum_{\bs{G}\,\in\Lambda^{\star}}
\left(
\tilde{f}^{1;\bs{G}}_{\bs{0}}
+
\tilde{f}^{2;\bs{G}}_{\bs{0}}
-
\sum_{\bs{q}}\;
U^{\,}_{\bs{q}}\,
\tilde{h}^{\bs{G}}_{-\bs{q}}
\right)
\hat{\varrho}^{\bs{G}}_{\bs{0}},
\\
&
\hat{H}^{\prime\bar{a}}_{U}=
\sum_{\bs{q}\in\Omega^{\,}_{\mathrm{BZ}}}\!
\sum_{\bs{G},\bs{G}'\,\in\Lambda^{\star}}\!
\tilde{f}^{4;\bs{G} }_{\bs{q}}
\tilde{f}^{4;-\bs{G}'}_{-\bs{q}}\,
e^{
-\mathrm{i}\,
(
\Theta^{\bs{G}}_{+\bs{q}}
+
\Theta^{-\bs{G}'}_{-\bs{q}}
)
  }
\,
\hat{\varrho}^{\bs{G} }_{\bs{q}}\,
\hat{\varrho}^{-\bs{G}'}_{-\bs{q}}.
\end{split}
\label{eq: final rep}
\end{equation}
Equation~(\ref{eq: MAIN RESULT})
follows.

\end{document}